%% file: main.tex
\documentclass[sigconf]{acmart}

\usepackage{algorithm}
\usepackage{algpseudocode} 
\usepackage{multirow}
\usepackage[normalem]{ulem}
\usepackage{subcaption} 
\usepackage{url}
\usepackage{dsfont}
\usepackage{amsmath}
\usepackage{caption}

\useunder{\uline}{\ul}{}

\AtBeginDocument{%
  }

\copyrightyear{2026}
\acmYear{2026}
\setcopyright{cc}
\setcctype{by}
\acmConference[MM '26]{Proceedings of the 34th ACM International Conference on Multimedia}{November 10--14, 2026}{Rio de Janeiro, Brazil}
\acmBooktitle{Proceedings of the 34th ACM International Conference on Multimedia (MM '26), November 10--14, 2026, Rio de Janeiro, Brazil}
\acmDOI{10.1145/3767308.3835045}
\acmISBN{979-8-4007-2213-4/2026/11}

\begin{document}

\title{SG-UMP: Sequence-Guided Universal Multimodal Prioritization
Calculation Framework}


\author{Xinyi Zhang}
\orcid{0009-0008-1950-3034}
\affiliation{%
  \institution{Imperial College London}
  \city{London}
  \country{UK}}
\email{zxyzxy090588@163.com}

\author{Yutong Li}
\orcid{0009-0008-6496-8850}
\affiliation{%
  \institution{University College London}
  \city{London}
  \country{UK}
}
\email{lyt3612671@163.com}

\author{Peijie Sun}
\authornote{Corresponding author.}
\orcid{0000-0001-9733-0521}
\affiliation{%
  \institution{Nanjing University of Posts and Telecommunications}
  \city{Nanjing}
  \country{China}}
\email{sun.hfut@gmail.com}

\renewcommand{\shortauthors}{Xinyi Zhang, Yutong Li and Peijie Sun}

\input{1-Abstract}
\input{2-Introduction}

\input{3-Preliminary}

\input{4-Methodology}
\input{5-Experiment}
\input{6-Related_Work}
\input{7-Conclusion}

\input{9-Acknowledge}

\bibliographystyle{ACM-Reference-Format}
\bibliography{MM26}

\input{8-Appendix}

\end{document}

%% file: 1-Abstract.tex
\begin{abstract}
Multimodal sequential recommendation (MSR) improves recommendation by incorporating heterogeneous information such as text, images, and user interactions. However, existing MSR methods often fail to capture user-level preference heterogeneity and dataset-level modality bias, limiting their adaptability across users and datasets. To address this issue, we propose \textbf{S}equence-\textbf{G}uided \textbf{U}niversal \textbf{M}ultimodal \textbf{P}rioritization Calculation Framework (\textbf{SG-UMP}), a plug-and-play plugin for enhancing multimodal information processing in MSR. SG-UMP includes a Module Combiner for flexible multimodal processing and a Module Router for dynamic module ordering, enabling adaptation to both user preferences and dataset characteristics. Experiments on four real-world datasets show that SG-UMP consistently improves recommendation performance across different backbones and multimodal settings. The code is available at ~\url{https://github.com/esemsc-xz524/SG-UMP}.
\end{abstract}

\ccsdesc[500]{Information systems~Recommender systems}

\keywords{Multimodal Recommendation, User Modeling, Dataset-aware Modeling, Module Routing}

\maketitle

%% file: 2-Introduction.tex
\section{Introduction}

Sequential recommendation (SR) has become indispensable in modern applications, aiming to mitigate information overload by predicting the next item a user is likely to interact with based on their historical behavior sequence~\cite{SASRec, SR_Survey1, SR_Survey2}.  With the rapid growth of multimodal content, incorporating multimodal information into SR has emerged as an effective solution to address the data sparsity and cold-start issues faced by traditional ID-based methods~\cite{TrustSVD}.  Multimodal sequential recommendation (MSR) integrates diverse sources of information, such as textual, visual, and acoustic contents, to capture user preferences and item properties more comprehensively, thereby improving recommendation accuracy~\cite{UniSRec, MMSR, MMMLP}.

\begin{figure}[ht!]
    \centering
    \includegraphics[width= \linewidth]{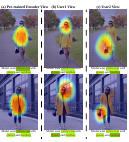}
    \caption{The visualization results on the Amazon dataset. Columns (a), (b), and (c) show the perspectives of the pre-trained encoder, User 1, and User 2, respectively. Red text indicates key attributes, and the intensity of the green background reflects attention assigned by the model.}
    \Description{The Visualization analysis results on the Amazon dataset.}
    \label{fig:intro}
\end{figure}

However, existing MSR methods typically rely on pre-trained encoders to extract multimodal embeddings. Since these encoders are generally trained on large-scale generic datasets such as ImageNet, the learned representations often fail to capture the modality-specific biases from users~\cite{Resnet, Bert, PreTrain_Noise1}. In practice, different users tend to attend to different aspects of multimodal information, resulting in distinct attention patterns over textual and visual contents. To illustrate this issue, Figure~\ref{fig:intro} visualizes multimodal attention on the Amazon dataset. Columns (a), (b), and (c) respectively show the generalized attention distribution of the pre-trained encoder and the approximated user-specific attention patterns of User 1 and User 2. As explicit user feedback or eye-tracking signals are unavailable in offline datasets, we approximate user attention based on historical interaction behaviors and salient item attributes, such as frequently interacted keywords and image regions highlighted by Class Activation Mapping (CAM)~\cite{CAM1, CAM2, CAM3}. As shown in the figure, User 1 pays more attention to accessories such as sunglasses, while User 2 focuses more on handbags. By contrast, the pre-trained encoder mainly attends to the coat, revealing its limited ability to adapt to user-specific interests. This mismatch indicates that generic multimodal representations derived from pre-trained encoders are insufficient for modeling heterogeneous user preferences~\cite{PreTrain_Noise1, MDSBR, BeFA}.

These observations reveal two limitations of existing MSR methods. (1) \textbf{User-level preference heterogeneity}: As discussed above, most existing MSR methods rely on pre-trained encoders to extract multimodal embeddings. Since these encoders are trained on large-scale generic datasets, the resulting representations are inherently generic and primarily capture common semantic patterns rather than user-specific interests. Consequently, the extracted multimodal representations remain insufficiently personalized, making it difficult to distinguish how users attend to different modalities and attributes. This limits current MSR models in accurately reflecting fine-grained user-specific interests in recommendation~\cite{PreTrain_Noise1, MDSBR, BeFA}. 
(2) \textbf{Dataset-level modality bias}: Existing MSR methods largely overlook dataset-level modality bias, where the relative informativeness of different modalities varies across datasets. For instance, in Amazon-style e-commerce scenarios, product images often serve as a primary source of user decision-making, as visual appearance, style, and design directly affect purchase preferences. By contrast, in Yelp-style local service platforms, textual information such as user reviews and business descriptions is often more informative, since user choices depend more on service quality, atmosphere, and experience than on visual presentation alone~\cite{VBPR, BeFA, Yelp1, Yelp2}. These dataset-dependent modality differences directly affect which signals are most beneficial for recommendation. Without the ability to adapt to such deviation, a model may over-rely on weaker modalities and underutilize the dominant ones, leading to suboptimal recommendation performance and poor cross-dataset generalization. Nevertheless, most existing MSR methods employ fixed multimodal processing pipelines, which are insufficiently flexible to accommodate dataset-level differences in modality importance~\cite{UniSRec, MMSR, MMMLP, MMSBR}.

Therefore, we argue that improving MSR requires adaptation at both user and dataset levels. On the one hand, the model should capture user-specific attention across modalities to better reflect personalized interests. On the other hand, it should adjust its processing strategy to the modality characteristics of different datasets. Accordingly, we derive two key design principles for an effective and flexible recommendation framework. \textbf{Principle 1}: Users exhibit different attention preferences across modalities, which can be modeled by leveraging their historical interaction sequences to guide personalized multimodal representation processing. \textbf{Principle 2}: Different datasets exhibit distinct modality distributions, dominant information sources, and data qualities. Accordingly, the framework should support modular and re-orderable processing to flexibly compose and arrange functional modules for each dataset.

Guided by the above principles, we propose \textbf{S}equence-\textbf{G}uided \textbf{U}niversal \textbf{M}ultimodal \textbf{P}rioritization Calculation Framework (\textbf{SG-UMP}), a plug-and-play plugin designed to enhance multimodal information processing in MSR. Rather than replacing the SR backbone itself, SG-UMP improves multimodal information processing in existing MSR models. To address user-level preference heterogeneity, SG-UMP introduces a \textbf{Module Combiner}, which flexibly integrates three representative functional modules for multimodal processing: filter-enhanced modules~\cite{FMLP, FFT1, FFT2} for reducing noise and extracting stable frequency-domain patterns, self-attention based modules~\cite{SASRec, Bert4rec} for adaptively fusing informative signals, and Mixture-of-Experts (MoE) based modules~\cite{MOE, PLE} for modeling interactions among multimodal features. These modules operate with distinct responsibilities and specialized functions, forming a comprehensive pipeline for multimodal embedding. They enable the model to capture heterogeneous user attention across modalities, thereby achieving better alignment with user-specific interests. To address dataset-level modality bias, SG-UMP incorporates a \textbf{Module Router}, which dynamically determines module execution order according to dataset-level modality characteristics and input-specific properties. In this way, SG-UMP adapts its multimodal processing flow to different modality distributions and data qualities, making the plugin more flexible and effective across recommendation scenarios.

We evaluate SG-UMP on four real-world datasets with diverse modality characteristics. Results show that each functional module consistently improves recommendation performance, while the optimal module order varies across datasets, confirming the importance of adaptive routing. Moreover, SG-UMP can be seamlessly integrated into different SR backbones, demonstrating its effectiveness, generality, and scalability. The main contributions are summarized as follows:

(1) We present two key design principles for MSR: personalized multimodal processing guided by user interaction sequences, and modular, re-orderable processing adaptable to different dataset characteristics.

(2) We propose SG-UMP, a plug-and-play plugin for enhancing multimodal information processing in MSR. It includes a Module Combiner for flexible multimodal processing and a Module Router for dynamic module ordering.

(3) We conduct extensive experiments on four real-world datasets, showing that SG-UMP consistently improves performance across different backbones and multimodal settings.

%% file: 3-Preliminary.tex
\section{Preliminary}

\subsection{Problem Definition}

MSR aims to exploit multimodal information and users’ historical behaviors to generate personalized recommendations for their next interactions. Given a set of users $\mathcal{U}$ and a set of items $\mathcal{Z}$, historical interactions can be chronologically organized into sequences. For each user $u \in \mathcal{U}$, we denote the interaction sequence as $\mathcal{S}^u = [x^u_1, x^u_2, \dots, x^u_{|\mathcal{S}^u|}]$, where each item $x^u_i \in \mathcal{I}$ denotes the item that the user interacted with at the $i$-th time step. Following most prior studies~\cite{MMMLP, MMSBR}, we focus on image and text modalities in our derivations, while the framework can be readily extended to accommodate additional modalities. Formally, each item is represented as $x_i = \{x_i^{id}, x_i^{img}, x_i^{txt}\}$, which incorporates the item ID, image, and textual description. The goal of MSR is to jointly model the sequential dependencies within $\mathcal{S}^u$ and the multimodal features of items to predict the next item a user will interact with.

\subsection{Multimodal Features Extraction}

For image $x_{i}^{img}$ and text $x_{i}^{txt}$, we adopt the CLIP encoder~\cite{CLIP} pretrained on large-scale corpora to obtain corresponding embeddings: 
${e}_{i}^{img} = \operatorname{CLIP}(x_{i}^{img})$, ${e}_{i}^{txt} = \operatorname{CLIP}(x_{i}^{txt})$.
The ID embedding layer $\operatorname{itemEmb}(\cdot)$is trainable to capture collaborative
signals in the interaction sequence:
${e}_{i}^{ID} = \operatorname{itemEmb}(x_{i}^{id})$.
To unify heterogeneous item modalities, each embedding is projected into a shared $d$-dimensional feature space for downstream fusion.

%% file: 4-Methodology.tex
\section{Methodology}

\label{setction: Method}
\begin{figure*}[htbp]
    \centering
    \includegraphics[width= \linewidth]{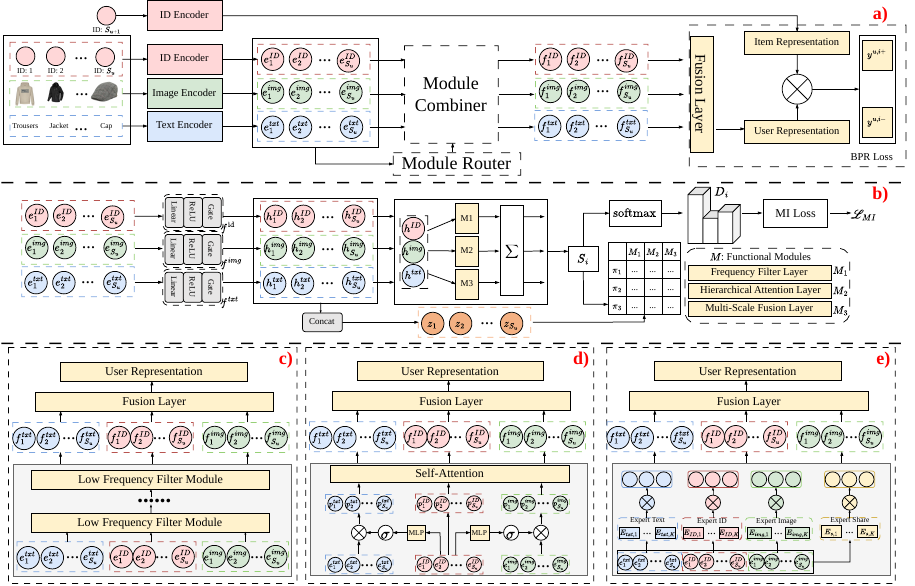}
    \caption{Illustration of SG-UMP and its components. (a) Overall framework of SG-UMP, where the Module Combiner integrates the Frequency Filter Layer, Hierarchical Attention Layer, and Multi-Scale Fusion Layer. (b) Module Router structure. (c) Frequency Filter Layer. (d) Hierarchical Attention Layer. (e) Multi-Scale Fusion Layer.}
    \Description{The Method Figure}
    \label{fig:method}
\end{figure*}

\subsection{Design Principle}

The SG-UMP is guided by two core principles designed to enhance multimodal adaptability to heterogeneous user attention and dataset modality variation. 
\textbf{Principle 1: User-Specific Multimodal Attention.}
Users exhibit heterogeneous preferences across different modalities. To be effective, MSR must account for these variations during representation learning and fusion. SG-UMP leverages historical interaction sequences as personalized signals to guide multimodal modeling, ensuring that information aligned with user interests is extracted or selectively emphasized.
\textbf{Principle 2: Modular and Re-orderable Modeling.} 
The informativeness and scale of modalities vary across datasets and domains, necessitating adaptive processing strategies. Consequently, the framework is designed to be modular and re-orderable, allowing various functional modules to be flexibly composed and dynamically sequenced based on input characteristics and dataset-level modality composition.

Guided by these principles, SG-UMP organizes multimodal modeling through two key components. (1) \textbf{Module Combiner} integrates three types of functional modules for representation learning: Frequency Filter Layer \(M_1\), Hierarchical Attention Layer \(M_2\), and Multi-Scale Fusion Layer \(M_3\). (2) \textbf{Module Router} determines their execution order in a dataset-driven manner. This architecture enables adaptive multimodal processing while maintaining flexibility and scalability across diverse recommendation scenarios.

\subsection{Frequency Filter Layer}

The Frequency Filter Layer extracts stable low-frequency patterns from multimodal sequences for robust representation learning. It captures user-specific preference patterns under Principle 1 while preserving modular consistency under Principle 2. Given multimodal embeddings of user interaction sequences, the layer transforms inputs $\mathbf{e}_i^{id}$, $\mathbf{e}_i^{txt}$, and $\mathbf{e}_i^{img}$ into refined outputs $\mathbf{f}_i^{id}$, $\mathbf{f}_i^{txt}$, and $\mathbf{f}_i^{img}$ with preserved dimensionality. The depth of the layer is controlled by a parameter $L$, enabling flexible adaptation to different tasks.

As illustrated in Figure~\ref{fig:method}(c), the layer operates in three steps: (1) transforming embeddings into the frequency domain; (2) filtering high-frequency components to capture stable low-frequency patterns; and (3) reconstructing refined embeddings in the time domain. Additional implementation details and theoretical analysis of the Frequency Filter Layer are provided in \textbf{Appendix A.1}. This design suppresses high-frequency noise while preserving stable low-frequency signals, producing informative and user-aligned multimodal representations.

\subsection{Hierarchical Attention Layer}

The Hierarchical Attention Layer models user-specific attention across modalities through adaptive weighting. It captures user-specific preference patterns under Principle 1 while preserving modular consistency under Principle 2. Given multimodal embeddings $\mathbf{e}_i^{id}$, $\mathbf{e}_i^{txt}$, and $\mathbf{e}_i^{img}$ derived from user interaction histories, the layer produces refined outputs $\mathbf{f}_i^{id}$, $\mathbf{f}_i^{txt}$, and $\mathbf{f}_i^{img}$ with preserved dimensionality for seamless integration with subsequent modules.

As illustrated in Figure~\ref{fig:method}(d), modality importance is measured by a Multi-Layer Perceptron (MLP) followed by a Sigmoid-based activation function:
\begin{align}
M(\mathbf{e}_i^{id}) &= \sigma(\operatorname{MLP}(\mathbf{W}\mathbf{e}_i^{id} + \mathbf{b})),
\end{align}
where $\mathbf{W}$ and $\mathbf{b}$ are trainable parameters, $\sigma$ denotes the Sigmoid activation, and $\operatorname{MLP}(\cdot)$ is a two-layer MLP. The user-interest-aware representations are then computed as:
\begin{equation}
\mathbf{p}_i^{m} = M(\mathbf{e}_i^{id}) \cdot \mathbf{e}_i^{m}, \quad m \in \{txt, img\},
\end{equation}
\begin{equation}
\mathbf{p}_i^{id} = \mathbf{e}_i^{id}.
\end{equation}
These representations are further refined using self-attention:
\begin{equation}
\mathbf{f}_i^{m} = \operatorname{Self-Attention}(\mathbf{p}_i^{m}), \quad m \in \{txt, img, id\}.
\end{equation}
This design adaptively emphasizes modality signals aligned with user preferences.

\subsection{Multi-Scale Fusion Layer}

The Multi-Scale Fusion Layer, illustrated in Figure~\ref{fig:method}(e), adopts the MoE framework for adaptive multimodal fusion. It is designed to capture user-specific attention across modalities under Principle 1 while preserving modularity and seamless integration under Principle 2. Specifically, the layer allocates computation between modality-unique experts and modality-shared experts, thereby supporting both fine-grained modality modeling and cross-modal interaction learning. For each modality \(m \in \{txt, img, id\}\), a set of modality-unique experts is constructed to extract distinctive features from the corresponding embedding \(e_i^m\), while an additional set of shared experts is introduced to model global patterns and dependencies across modalities. The numbers of modality-unique experts \(U\) and modality-shared experts \(S\) are tunable hyperparameters, allowing the layer to adapt to datasets with different modality relevance and complexity.

To dynamically control expert contributions, we employ a lightweight gating network. For each modality, the gating score is computed as:
\begin{align}
G_{m} &= \operatorname{softmax}(W_{g2} \, \sigma(W_{g1} e_i^m + b_{g1}) + b_{g2}), \quad m \in \{txt, img, id\},
\end{align}
where \(W_{g1} \in \mathbb{R}^{d \times h}\) and \(W_{g2} \in \mathbb{R}^{h \times K}\) are learnable projection matrices, \(h\) is the hidden dimension, \(\sigma(\cdot)\) is a nonlinear activation function, and \(\operatorname{softmax}(\cdot)\) denotes softmax normalization that converts raw scores into gating probabilities. Based on these scores, the modality-unique representation is obtained by
\begin{align}
f_i^m = \sum_{j=1}^{U} G_{m,j} \cdot E_{m,j}(e_i^m), \quad m \in \{txt, img, id\},
\end{align}
where \(G_{m,j}\) is the gating score and \(E_{m,j}(\cdot)\) is the output of the \(j\)-th modality-unique expert for modality \(m\). 

For modality-shared experts, the concatenated multimodal embedding
\(e_i = \operatorname{concat}(e_i^{img}, e_i^{txt}, e_i^{id})\) is processed in the same way:
\begin{align}
G_{s} &= \operatorname{softmax}(W_{g4} \, \sigma(W_{g3} e_i + b_{g3}) + b_{g4}), \\
f_i^s &= \sum_{j=1}^{S} G_{s,j} \cdot E_{s,j}(e_i).
\end{align}
By combining modality-unique and modality-shared representations, the Multi-Scale Fusion Layer produces comprehensive multimodal embeddings that preserve both modality uniqueness and representation consistency across modalities.

\subsection{Module Router}

The Module Router dynamically determines the execution order of functional modules to facilitate adaptive multimodal processing. Given multimodal embeddings, the router computes dispatch scores and derives routing permutations to select module sequences that best match input characteristics and dataset-level modality composition, as shown in Figure~\ref{fig:method}(b). To promote input-aware and stable routing, the router is regularized by a conditional mutual information objective, which strengthens the dependency between modalities and modules while encouraging diverse routing behaviors. Let $\mathcal{D} = \{D_1, D_2, \ldots, D_n\}$ denote a set of domain-specific datasets, where each dataset contains user–item interaction sequences comprising ID, image, and text embeddings. Given that modality informativeness and contributions vary across domains, the router learns a dataset-adaptive execution order for the three functional modules within the Module Combiner.

\subsubsection{Dynamic Module Routing}
To dynamically determine the execution order of three functional modules, we introduce a modality-aware router that analyzes the multimodal composition of each input and assigns execution priorities accordingly. Given an input representation \(\mathbf{e}_i = \{\mathbf{e}_i^{id}, \mathbf{e}_i^{img}, \mathbf{e}_i^{txt}\}\), each embedding is first processed by a dedicated gating network:
\begin{equation}
\mathbf{h}_i^{m} = \operatorname{f^m}(\mathbf{e}_i^{m}), \quad m \in \{id, img, txt\},
\end{equation}
where $\operatorname{f^m(\cdot)}$ denotes a modality-specific gating function composed of a linear projection, a non-linear activation, and a gating mechanism that regulates information flow. 
This design allows independent yet controlled processing of each modality, ensuring that salient modality-specific cues are preserved for subsequent routing. For each functional module \( M_k \in \{M_1, M_2, M_3\} \) and each modality \( m \), the router computes a dispatch score:
\begin{equation}
\mathbf{s}_{i,k}^{m} = \mathbf{w}_k^{\top} \cdot \mathrm{ReLU}(\mathbf{W}^{m}\mathbf{h}_i^{m} + \mathbf{b}_k), \quad m \in \{id, img, txt\}, \quad k \in \{1, 2, 3\},
\end{equation}
where \( \mathbf{W}^{m} \in \mathbb{R}^{d \times d} \), \( \mathbf{w}_k \in \mathbb{R}^{d} \), and \( \mathbf{b}_k \in \mathbb{R}^{d} \) are learnable parameters. 
Scores from all modalities are aggregated into a unified module preference for each mudule and normalized by softmax to produce the routing distribution:
\begin{equation}
\mathbf{s}_{i,k} = \sum_{m} \mathbf{s}_{i,k}^{m}, \quad m \in \{id, img, txt\}, \quad k \in \{1, 2, 3\},
\end{equation}
\begin{equation}
\mathbf{D}_i = \mathrm{softmax}(\mathbf{s}_{i,k}), \quad k \in \{1, 2, 3\}.
\end{equation}

Based on the dispatch scores, the router determines the optimal execution order of modules by sorting \(\mathbf{D}_{i,k}\) in descending order, yielding a routing permutation \(\pi = (\pi_1, \pi_2, \pi_3)\), where each \(\pi_j\) indicates the module index executed at step \(j\). 
Finally, the fused multimodal representation is sequentially processed according to the routing order:
\begin{equation}
\mathbf{z}_i = [\mathbf{h}_i^{id}; \mathbf{h}_i^{img}; \mathbf{h}_i^{txt}], \quad 
\mathbf{y}_i = M_{\pi_3}\!\left(M_{\pi_2}\!\left(M_{\pi_1}(\mathbf{z}_i)\right)\right).
\end{equation}

\subsubsection{Conditional Mutual Information Loss}
To establish explicit dependencies between modalities and functional modules, we introduce a regularization term based on conditional mutual information (CMI). This loss encourages input-dependent routing and prevents uniform or degenerate execution order of modules. 
Let $K$ and $N$ denote the numbers of modules and modalities, respectively, with $m \in \{id, img, txt\}$ and $M_k$ the $k$-th module. Each input $\mathbf{e}_i = \{\mathbf{e}_i^{id}, \mathbf{e}_i^{img}, \mathbf{e}_i^{txt}\}$ consists of modality-specific embedding $\mathbf{e}_i^m$, which produce gating outputs $\mathbf{h}_i^m$ and dispatch scores $\mathbf{s}_{i,k}^m$. The soft routing vector $\mathbf{D}_i \in \mathbb{R}^{K}$ is interpreted as the conditional distribution $P(M_k \mid \mathbf{e}_i)$. 
The joint probability of assigning modality $m$ to module $M_k$ is approximated as:
\begin{equation}
P(M_k, m \mid \mathbf{e}_i) = 
\frac{\mathbf{s}_{i,k}^{m}}{\sum_{k=1}^{K} \sum_{m=1}^{N} \mathbf{s}_{i,k}^{m}}.
\end{equation}
The corresponding marginal and conditional probabilities are defined as:
\begin{equation}
P(m \mid \mathbf{e}_i) = 
\frac{\sum_{k=1}^{K} \mathbf{s}_{i,k}^{m}}{\sum_{k=1}^{K} \sum_{m=1}^{N} \mathbf{s}_{i,k}^{m}}, 
\quad
P(M_k \mid \mathbf{e}_i) = 
\frac{\sum_{m=1}^{N} \mathbf{s}_{i,k}^{m}}{\sum_{k=1}^{K} \sum_{m=1}^{N} \mathbf{s}_{i,k}^{m}}.
\end{equation}

Based on these probabilities, we define the conditional mutual information loss as:
\begin{equation}
\begin{aligned}
\mathcal{L}_{MI} 
= - I(M; m \mid \mathbf{e}_i)
= - \sum_{k=1}^{K} \sum_{m=1}^{N} 
P(M_k, m \mid \mathbf{e}_i)\\
\log \frac{P(M_k, m \mid \mathbf{e}_i)}
{P(M_k \mid \mathbf{e}_i) P(m \mid \mathbf{e}_i)}.
\end{aligned}
\end{equation}

By maximizing the conditional mutual information, the router is encouraged to align modality cues with the behaviors of corresponding modules. For example, image-dominant inputs are directed to image-processing modules, whereas text-dominant inputs are routed to text modules. This regularization promotes diverse and input-sensitive routing strategies, ensuring modality-aware execution and preventing degenerate cases where all inputs follow identical execution paths. A detailed theoretical analysis of this objective is provided in \textbf{Appendix A.2}.

\subsection{Training and Prediction}
\subsubsection{Module Processing and Fusion Layer.}
The framework processes multimodal embeddings of image, text, and ID through three functional modules, including the Frequency Filter, Self-Attention, and MoE layers. The Module Router dynamically determines the execution order based on dataset-level input characteristics. Each module produces representations of consistent dimensionality, which are aggregated via mean pooling to obtain the final user representation, capturing both modality-unique and shared information.
\subsubsection{Training Objective.}
The model is trained with the Bayesian Personalized Ranking (BPR) loss~\cite{MFBPR} to optimize implicit feedback by maximizing preference scores of positive over negative samples. To encourage input-aware and diverse routing, a conditional mutual information regularization term is introduced. The overall loss is:
\begin{equation}
\mathcal{L} = \mathcal{L}_{\text{BPR}} + \alpha \mathcal{L}_{MI},
\end{equation}
where $\mathcal{L}_{\text{BPR}}$ denotes the BPR loss, $\mathcal{L}_{MI}$ is the mutual information regularization, and $\alpha$ is a hyperparameter. The detailed training procedure is provided in \textbf{Appendix A.3}.

%% file: 5-Experiment.tex
\section{Experiments}
To comprehensively evaluate the performance and design principles of SG-UMP, we conducted experiments focusing on the following research questions:
\textbf{RQ1:} How effective are Principles 1 and 2 in enhancing the performance of SG-UMP and adaptability in multimodal recommendation tasks?
\textbf{RQ2:} How does SG-UMP perform across various sequential recommendation models?
\textbf{RQ3:} How do different hyperparameter settings influence the performance of SG-UMP? 
\textbf{RQ4:} What is the time complexity of SG-UMP?

\subsection{Experimental Setup}
\textbf{Datasets.} We evaluate our model on four datasets: Amazon\footnote{\url{https://snap.stanford.edu/data/amazon/productGraph}} Home, Beauty, Office, and Yelp\footnote{\url{https://business.yelp.com/data/resources/open-dataset}}. The Amazon datasets capture purchasing behavior and reviews across diverse domains, serving as widely used benchmarks for SR~\cite{SASRec, STOSA}. The Yelp dataset, sourced from a business platform, contains restaurant reviews and is widely used in recommendation tasks~\cite{Yelp1, Yelp2}. All datasets are pre-processed using a 5-core setting~\cite{SASRec, STOSA, Yelp1, Yelp2}, ensuring each user and item has at least five interactions. Detailed dataset statistics are provided in Table~\ref{tab:Statistics}.

\begin{table}[h!]
\centering
\caption{Datasets Statistics}
\vspace{-0.3cm}
\resizebox{0.45\textwidth}{!}{%
\begin{tabular}{@{}cccccc@{}}
\toprule
Dataset & \multicolumn{1}{c}{\#users} & \multicolumn{1}{c}{\#items} & \multicolumn{1}{c}{\#interactions} & \multicolumn{1}{c}{density} & \multicolumn{1}{c} {avg.length} \\ 
\midrule
Home & 66,519 & 28,237 & 551,682 & 0.03\% & 8.3 \\
Beauty & 22,363 & 12,101 & 198,502 & 0.05\% & 8.3 \\
Office& 4,905& 2,421& 53,258& 0.04\%& 10.8\\
Yelp & 287,116 & 148,523 & 4,392,169 & 0.01\% & 15.3\\
\bottomrule
\end{tabular}%
}
\label{tab:Statistics}
\end{table}

\textbf{Evaluation Settings.} To ensure fair and rigorous evaluation, we adopt the standard leave-one-out strategy~\cite{SASRec, STOSA}, where the last item in each user sequence is used for testing, the second-to-last for validation, and the rest for training. We report performance using widely-used metrics~\cite{SASRec, MISSRec, PreTrain_Noise1, FMLP}: Recall@K (R@K) and NDCG@K (N@K), with $K = 10$ and $20$. Higher R@K and N@K indicate better recommendation accuracy.

\textbf{Baselines.} We compare our proposed model with representative recommendation models, categorized into two groups: 
\textbf{(1) Traditional Recommendation:} SASRec~\cite{SASRec}, BERT4-Rec~\cite{Bert4rec}, LightGCN~\cite{LightGCN}, STOSA~\cite{STOSA}, FMLP-Rec~\cite{FMLP} and Oracle4Rec~\cite{Oracle4Rec}. \textbf{(2) Multimodal Recommendation:}  
VBPR~\cite{VBPR}, MMMLP~\cite{MMMLP}, MML~\cite{MML}, MMSR~\cite{MMSR}, MMSBR~\cite{MMSBR} and MP4SR~\cite{MP4SR}.  

\textbf{Implementation Details.}
For fair comparison, baselines are configured using reported hyperparameters or tuned via grid search on validation sets when unavailable. We adopt CLIP ViT-B/32\footnote{\url{https://huggingface.co/sentence-transformers/clip-ViT-B-32}} as the unified encoder for both text and image modalities. Key hyperparameters are selected via grid search, including the number of shared and task experts $\{1,2,3,4,5\}$, the mutual information loss weight $\alpha \in \{0.1,0.2,0.5,1.0,2.0\}$, and the number of Frequency Filter layers $\{1,2,3,4,5\}$. Results are averaged over 5 runs and are statistically significant ($p<0.05$). All experiments are conducted on an NVIDIA 4090 GPU.

\begin{table*}[!ht]
\caption{Validation of Module Importance and Sequence Effect on Model Performance. The best performance is highlighted in bold. The table has two parts: (1) The importance and impact of different modules. \underline{$w/o$ Filter}: Excluding the Frequency Filter Layer. \underline{$w/o$ Attention}: Excluding Hierarchical Attention Layer. \underline{$w/o$ MOE}: Excluding the Multi-Scale Fusion Layer. (2) The effect of different module sequence combinations. (a) is Frequency Filter Layer, (b) is Hierarchical Attention Layer, and (c) is Multi-Scale Fusion Layer. All results are averaged over 5 runs per dataset to ensure statistical robustness and are statistically significant with \(p < 0.05\).}
\vspace{-0.3cm}
\centering
\resizebox{0.96\textwidth}{!}{%
\begin{tabular}{cc ccccc | cc cccccc}
\toprule
\multicolumn{7}{c|}{(1) The Importance and Effect of different Module.} & \multicolumn{8}{c}{(2) The Importance and Effect of different Module Sequence.} \\\hline
Datasets& Metric & STOSA & $w/o$ Filter & $w/o$ Attention & $w/o$ MOE & SG-UMP+STOSA & Datasets& Metrics & a-b-c  & a-c-b & b-a-c & b-c-a & c-a-b & c-b-a \\\hline
\multirow{4}{*}{Home} 
& R@10  & 0.0169 & 0.0216 & 0.0213 & 0.0227 & \textbf{0.0242}& \multirow{4}{*}{Home}   & R@10 & 0.0239 & 0.0227 & 0.0229 & 0.0214 & 0.0229 & \textbf{0.0242}\\
& R@20  & 0.0264 & 0.0316 & 0.0314 & 0.0322 & \textbf{0.0349}&                         & R@20 & 0.0333 & 0.0328 & 0.0329 & 0.0319 & 0.0333 & \textbf{0.0349}\\
& N@10  & 0.0098 & 0.0121 & 0.0120 & 0.0124 & \textbf{0.0144}&                         & N@10 & 0.0121 & 0.0138 & 0.0133 & 0.0140 & 0.0131 & \textbf{0.0144}\\
& N@20  & 0.0113 & 0.0134 & 0.0132 & 0.0141 & \textbf{0.0167}&                         & N@20 & 0.0162 & 0.0144 & 0.0158 & 0.0146 & 0.0156 & \textbf{0.0167}\\\hline

\multirow{4}{*}{Beauty} 
& R@10  & 0.0648 & 0.0726 & 0.0723 & 0.0725 & \textbf{0.0788}& \multirow{4}{*}{Beauty} & R@10 & 0.0779 & 0.0774 & \textbf{0.0788} & 0.0780 & 0.0785 & 0.0766\\ 
& R@20  & 0.0941 & 0.1028 & 0.1021 & 0.1024 & \textbf{0.1091}&                         & R@20 & 0.1066 & 0.1073 & \textbf{0.1091} & 0.1083 & 0.1067 & 0.1026\\ 
& N@10  & 0.0339 & 0.0406 & 0.0408 & 0.0403 & \textbf{0.0455}&                         & N@10 & 0.0436 & 0.0449 & \textbf{0.0455} & 0.0443 & 0.0436 & 0.0446\\ 
& N@20  & 0.0385 & 0.0431 & 0.0433 & 0.0434 & \textbf{0.0534}&                         & N@20 & 0.0525 & 0.0524 & \textbf{0.0534} & 0.0512 & 0.0517 & 0.0532\\\hline 

\multirow{4}{*}{Office} 
& R@10  & 0.0904 & 0.0926 & 0.0923 & 0.0915 & \textbf{0.0975}& \multirow{4}{*}{Office} & R@10 & 0.0889 & \textbf{0.0975} & 0.0913 & 0.0922 & 0.0869 & 0.0922\\
& R@20  & 0.1388 & 0.1376 & 0.1382 & 0.1374 & \textbf{0.1454}&                         & R@20 & 0.1372 & \textbf{0.1454} & 0.1319 & 0.1425 & 0.1295 & 0.1394\\
& N@10  & 0.0516 & 0.0521 & 0.0520 & 0.0523 & \textbf{0.0563}&                         & N@10 & 0.0502 & \textbf{0.0563} & 0.0516 & 0.0535 & 0.0499 & 0.0527\\
& N@20  & 0.0626 & 0.0621 & 0.0623 & 0.0618 & \textbf{0.0683}&                         & N@20 & 0.0622 & \textbf{0.0683} & 0.0619 & 0.0661 & 0.0606 & 0.0645\\\hline

\multirow{4}{*}{Yelp} 
& R@10  & 0.0248 & 0.0256 & 0.0253 & 0.0255 & \textbf{0.0279}& \multirow{4}{*}{Yelp}   & R@10 & \textbf{0.0266} & 0.0256 & 0.0244 & 0.0233 & 0.0263 & 0.0245\\
& R@20  & 0.0424 & 0.0446 & 0.0444 & 0.0441 & \textbf{0.0492}&                         & R@20 & \textbf{0.0442} & 0.0433 & 0.0407 & 0.0391 & 0.0439 & 0.0405\\
& N@10  & 0.0128 & 0.0134 & 0.0136 & 0.0131 & \textbf{0.0151}&                         & N@10 & \textbf{0.0132} & 0.0128 & 0.0121 & 0.0114 & 0.0129 & 0.0120\\
& N@20  & 0.0161 & 0.0167 & 0.0165 & 0.0165 & \textbf{0.0184}&                         & N@20 & \textbf{0.0177} & 0.0172 & 0.0162 & 0.0154 & 0.0173 & 0.0160\\\hline
\end{tabular}
}
\label{tab:preciple1&2}
\end{table*}

\subsection{Effectiveness of Principle 1 and Principle 2 (RQ1)}

\textbf{Verification of Principle 1.}
To validate \textbf{Principle 1}, we conduct an ablation study on STOSA~\cite{STOSA} by removing each core module of SG-UMP: Frequency Filter Layer (\textit{w/o Filter}), Hierarchical Attention Layer (\textit{w/o Attention}), and Multi-Scale Fusion Layer (\textit{w/o MOE}). As shown in Table~\ref{tab:preciple1&2} (1), removing any module consistently degrades performance, showing that all components provide complementary signals. Specifically, removing the Frequency Filter Layer leads to a consistent drop across datasets, highlighting its role in denoising multimodal inputs. Removing the Hierarchical Attention Layer causes the largest degradation, especially on Beauty and Office, indicating the importance of modeling fine-grained user preferences. Removing the Multi-Scale Fusion Layer also results in notable performance loss, confirming the necessity of adaptive fusion for capturing shared and modality-specific information. Overall, the full SG-UMP achieves the best performance across all datasets, verifying that effective MSR requires both user-aware attention modeling and complementary module interactions.

\textbf{Verification of Principle 2.}
To validate \textbf{Principle 2}, we evaluate all six permutations of the three core modules: (a) Frequency Filter, (b) Hierarchical Attention, and (c) Multi-Scale Fusion. As shown in Table~\ref{tab:preciple1&2} (2), the optimal execution order varies across datasets (e.g., c-b-a for Home, b-a-c for Beauty, a-c-b for Office, and a-b-c for Yelp), indicating that no fixed pipeline is universally optimal. This confirms that module effectiveness depends on dataset-specific modality distributions and interaction patterns. Notably, datasets with longer sequences and higher sparsity (e.g., Yelp) benefit from early filtering and attention refinement, while denser datasets (e.g., Beauty and Office) favor earlier attention or fusion. These results show that the Module Router can adaptively select execution paths aligned with data characteristics, leading to consistent performance gains. Furthermore, recall-based metrics are more sensitive to module order than NDCG, indicating that different evaluation metrics respond differently to execution order.

\textbf{Discussion.}
We further analyze model behavior via attention visualization (Section~\ref{section:visual}). While these visualizations provide qualitative insights into user focus patterns, they do not reliably reflect the effectiveness of different modules. Therefore, we rely primarily on quantitative results for evaluation. Future work will explore more informative visualization and interpretability techniques to better understand how module interactions influence multimodal representation learning.

\begin{table*}[!ht]
    \small
    \setlength\tabcolsep{2pt}
    \vspace{-0.3cm}
    \caption{Performance comparison with two groups of baselines. The best overall performance is highlighted in bold, while the strongest baseline is marked by underlines. \emph{vs. Vanilla} represents the relative improvements over the vanilla model, and \emph{vs. Best} represents the relative improvements over the best baseline in percentage. All results are averaged over 5 runs per dataset to ensure statistical robustness and are statistically significant with \(p < 0.05\).}
    \label{tab:compare}
    \vspace{-5pt}
    \resizebox{0.96\linewidth}{!}{
        \begin{tabular}{c|cccc|cccc|cccc|cccc}
        \hline
        \makebox[0.15\linewidth]{Datasets}        & \multicolumn{4}{c|}{Home}                                                         & \multicolumn{4}{c|}{Beauty}                                                                                           & \multicolumn{4}{c|}{Office}                                                                                          & \multicolumn{4}{c}{Yelp}                                                                             \\ \hline
        Metric                                    & {R@10}          & {R@20}           & {N@10}          &{N@20}                      & {R@10}                      & {R@20}                      & {N@10}                      & {N@20}                      & {R@10}                       & {R@20}                     & {N@10}                    & {N@20}                       & {R@10}                            & {R@20}                          & {N@10}               & {N@20}  \\ \hline
        SASRec (ICDM'18)                          & 0.0168          & 0.0249           & 0.0081          & 0.0099                     & 0.0534                      & 0.0839                      & 0.0247                      & 0.0332                      & 0.0853& 0.1256& 0.0487& 0.0592& 0.0233                            & 0.0391                          & 0.0123               & 0.0152  \\
        BERT4Rec (CIKM'19)                        & 0.0156          & 0.0238           & 0.0073          & 0.0101                     & 0.0545                      & 0.0852                      & 0.0254                      & 0.0351                      & 0.0857& 0.1232& 0.0462& 0.0599& 0.0245                            & 0.0411                          & 0.0119               & 0.0162  \\ 
        LightGCN (SIGIR'20)                       & 0.0161          & 0.0243           & 0.0077          & 0.0103                     & 0.0549                      & 0.0861                      & 0.0258                      & 0.0355                      & 0.0866& 0.1215& 0.0463& 0.0602& 0.0236                            & 0.0414                          & 0.0123               & 0.0166  \\
        STOSA (WWW'22)                            & 0.0169          & 0.0264           & 0.0098          & 0.0113                     & 0.0648                      & 0.0941                      & 0.0339                      & 0.0385                      & 0.0864& 0.1284& 0.0497& 0.0602& 0.0238                            & 0.0424                          & 0.0128               & 0.0161  \\
        FMLPRec (WWW'22)                          & 0.0178          & 0.0268           & 0.0085          & 0.0108                     & 0.0654                      & 0.0961                      & 0.0348                      & 0.0381                      & 0.0881& 0.1321& 0.0488& 0.0601& 0.0247                            & 0.0438                          & 0.0131               & 0.0167  \\
        Oracle4Rec (WSDM'25)                          & 0.0211          & 0.0302           & 0.0131          & 0.0144                     & 0.0686                      & 0.0997                      & 0.0379                      & 0.0451                      & {\ul 0.0901}& {\ul 0.1419}& {\ul 0.0511}& 0.0610& 0.0266                            & 0.0445                          & 0.0146               & 0.0175  \\ \hline
        VBPR (AAAI'16)                            & 0.0159          & 0.0256           & 0.0074          & 0.0095                     & 0.0551                      & 0.0842                      & 0.0252                      & 0.0337                      & 0.0821& 0.1212& 0.0459& 0.0565& 0.0237                            & 0.0434                          & 0.0123               & 0.0163  \\
        MMMLP (WWW'23)                            & 0.0209          & 0.0293           & 0.0124          & 0.0134                     & 0.0671                      & 0.0979                      & 0.0387                      & 0.0441                      & 0.0872& 0.1291& 0.0501& 0.0603& 0.0254                            & 0.0452                          & 0.0138               & 0.0169  \\
        MML (CIKM'23)                             & {\ul 0.0237}    & 0.0327           & {\ul 0.0145}          & 0.0149                     & {\ul 0.0735}                & 0.1076                      & {\ul 0.0433}                                   & {\ul 0.0512}                &  0.0881& 0.1317& 0.0504& {\ul 0.0611}& {\ul 0.0291}                        & 0.0472                          & 0.0146             & 0.0171   \\
        MMSR (CIKM'23)                            & 0.0218          & 0.0322           & 0.0135          & 0.0142                     & 0.0715                      & 0.1062                      & 0.0401                      & 0.0446                      & 0.0878& 0.1367& 0.0505& 0.0601& 0.0245                            & 0.0456                          & 0.0129               & 0.0156  \\
        MMSBR (TKDE'23)                           & 0.0235          & 0.0325           & 0.0134          & {\ul 0.0158}               & 0.0727                      & 0.1088                      & 0.0427                      & 0.0504                      & 0.0865&  0.1325& 0.0503& 0.0599& 0.0271                            & 0.0464                          & {\ul 0.0151}         & 0.0181  \\
        MP4SR (TOIS'24)                           & 0.0223          & {\ul 0.0327}     & 0.0129          & 0.0155                     & 0.0731                      & {\ul 0.1092}              & 0.0425                      & 0.0511                      & 0.0898& 0.1321& 0.0509& 0.0604& 0.0281                            & {\ul 0.0483}                    & 0.0149               & {\ul 0.0186}  \\ \hline
        SG-UMP+SASRec                    & 0.0211          & 0.0299           & 0.0114          & 0.0141                     & 0.0688                      & 0.1043                      & 0.0356                      & 0.0443                      & 0.0960                       & 0.1423                     & 0.0527                    & 0.0640                       & 0.0267                            & 0.0435                          & 0.0137               & 0.0171  \\
        \emph{vs. Vanilla}              & 25.60\%         & 20.08\%          & 40.74\%         & 42.42\%                    & 28.84\%                     & 24.31\%                     & 44.13\%                     & 33.43\%                     & 12.54\%                      & 13.30\%                  & 8.21\%                 & 8.11\%                    & {14.59\%}                         & 11.25\%                         & 11.38\%              & 12.50\% \\
        SG-UMP+STOSA                     & 0.0242          & 0.0349           & 0.0144          & 0.0167                     & 0.0788                      & 0.1091                      & 0.0455                      & 0.0534                      & 0.0975                       & 0.1454                     & 0.0563                    & 0.0683                       & 0.0279                            & 0.0492                          & 0.0151               & 0.0184  \\
        \emph{vs. Vanilla}              & 43.20\%         & 32.20\%          & 46.94\%         & 47.79\%                    & 21.60\%                     & 15.94\%                     & 34.22\%                     & 38.70\%                     & 12.85\%                      & 13.24\%                  & 13.28\%                 & 13.46\%                    & 17.23\%                           & 16.04\%                         & 17.97\%              & 14.29\%     \\
        SG-UMP+Oracle4Rec                    & \textbf{0.0264} & \textbf{0.0361}  & \textbf{0.0165} & \textbf{0.0178}            & \textbf{0.0807}             & \textbf{0.1188}             & \textbf{0.0476}             & \textbf{0.0565}             & \textbf{0.1017}              & \textbf{0.1609}            & \textbf{0.0601}           & \textbf{0.0707}              & \textbf{0.0312}                   & \textbf{0.0516}                 & \textbf{0.0162}      & \textbf{0.0202}  \\
        \emph{vs. Vanilla}              & 25.12\%         & 20.86\%          & 25.95\%         & 23.61\%                    & 17.64\%                     & 19.16\%                     & 25.59\%                     & 25.28\%                     & 12.87\%                      & 13.39\%                  & 17.61\%                 & 15.90\%                    & 17.29\%                           & 15.96\%                         & 10.96\%              & 15.43\% \\ \hline
        \emph{vs. Best}                 & 11.39\%         & 10.40\%          & 13.79\%         & {12.66\%}                  & {9.80\%}                    & {8.79\%}                    & {9.93\%}                    & {10.35\%}                   & 12.87\%                     & 13.39\%                   & 17.61\%                  & 15.71\%                     & 7.22\%                            & 6.83\%                          & 7.28\%               & 8.60\% \\ \hline
        \end{tabular}
        }
    \vspace{-8pt}
\end{table*}

\subsection{Overall Comparison (RQ2)}

To evaluate the effectiveness and flexibility of SG-UMP, we integrate it into three representative SR backbones, including SASRec~\cite{SASRec}, STOSA~\cite{STOSA}, and Oracle4Rec~\cite{Oracle4Rec}, and conduct experiments on four datasets. These models represent different modeling paradigms. SASRec captures sequential dependencies via self-attention, STOSA introduces stochasticity through Wasserstein self-attention, and Oracle4Rec incorporates item frequency priors to enhance sequence modeling. This diversity provides a comprehensive benchmark for evaluating the robustness and adaptability of SG-UMP. As shown in Table~\ref{tab:compare}, SG-UMP consistently outperforms all baselines across datasets. On average, it improves SASRec by $21.96\%$, STOSA by $24.93\%$, and Oracle4Rec by $18.91\%$, demonstrating its effectiveness across different architectures. The results reveal three key observations: (1) SG-UMP significantly improves performance by incorporating multimodal information and aligning it with user preferences, enabling more accurate modeling of user behavior. (2) SG-UMP consistently outperforms existing multimodal methods because its modules do not simply fuse multimodal features, but process them in a user-aware manner. In particular, filtering suppresses irrelevant signals, while attention modeling better captures user-specific preferences across modalities. (3) SG-UMP remains effective across different backbones and datasets because its modular design and dynamic routing allow the processing flow to adapt to dataset-specific modality characteristics.

\subsection{Visualization Analysis (RQ2)}\label{section:visual}
\begin{figure}[htbp]  
    \centering  
    \includegraphics[width= \linewidth]{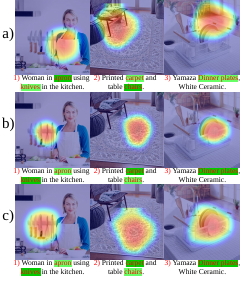}  
    \vspace{-0.9cm}
    \caption{Visualization analysis. (a), (b), and (c) correspond to STOSA with pre-trained embeddings, the approximated user perspective based on salient item features, and SG-UMP + STOSA, respectively. Red text highlights important attributes, and the intensity of the green background indicates attention assigned by the model.}
    \Description{The Visualization Analysis.}
    \label{fig:case-study}  
\end{figure}  

To demonstrate the effectiveness of SG-UMP in refining multimodal attention, we analyze attention maps under three settings: (a) STOSA using embeddings extracted by pre-trained encoders for multimodal feature extraction, (b) an approximated user perspective constructed from item features related to user preferences, and (c) the combined perspective of SG-UMP and STOSA. In setting (b), since offline datasets do not provide direct user feedback or precise attention records, we approximate user focus using item metadata and historical interaction signals. Salient item attributes, such as frequently interacted keywords, tags, or core descriptive features identified in prior studies, are selected according to their statistical relevance to user behavior. These attributes serve as proxies for aspects likely to attract user interest. For visual features, we highlight the corresponding regions using Class Activation Mapping (CAM)~\cite{CAM1, CAM2, CAM3}. For textual features, we identify key words in item descriptions that are aligned with these salient attributes. This design enables a comparison between model-generated attention and empirically grounded user-relevant features without requiring manual annotations. As shown in Figure~\ref{fig:case-study}(a), STOSA without user-specific modeling exhibits diffuse attention and often misses key product characteristics in both image and text. Figure~\ref{fig:case-study}(b) presents the constructed user reference, emphasizing attributes statistically associated with user preferences. Figure~\ref{fig:case-study}(c) shows that SG-UMP produces more focused attention aligned with these meaningful features, resulting in more personalized item representations. This analysis confirms that SG-UMP improves attention alignment with user-relevant signals and enhances the interpretability and relevance of multimodal recommendations.

\subsection{Parameter Sensitivity (RQ3)}
\begin{figure}[ht]
    \centering
    \label{fig:Hyper}

    \begin{subfigure}{0.22\textwidth}
        \centering
        \includegraphics[width=\linewidth]{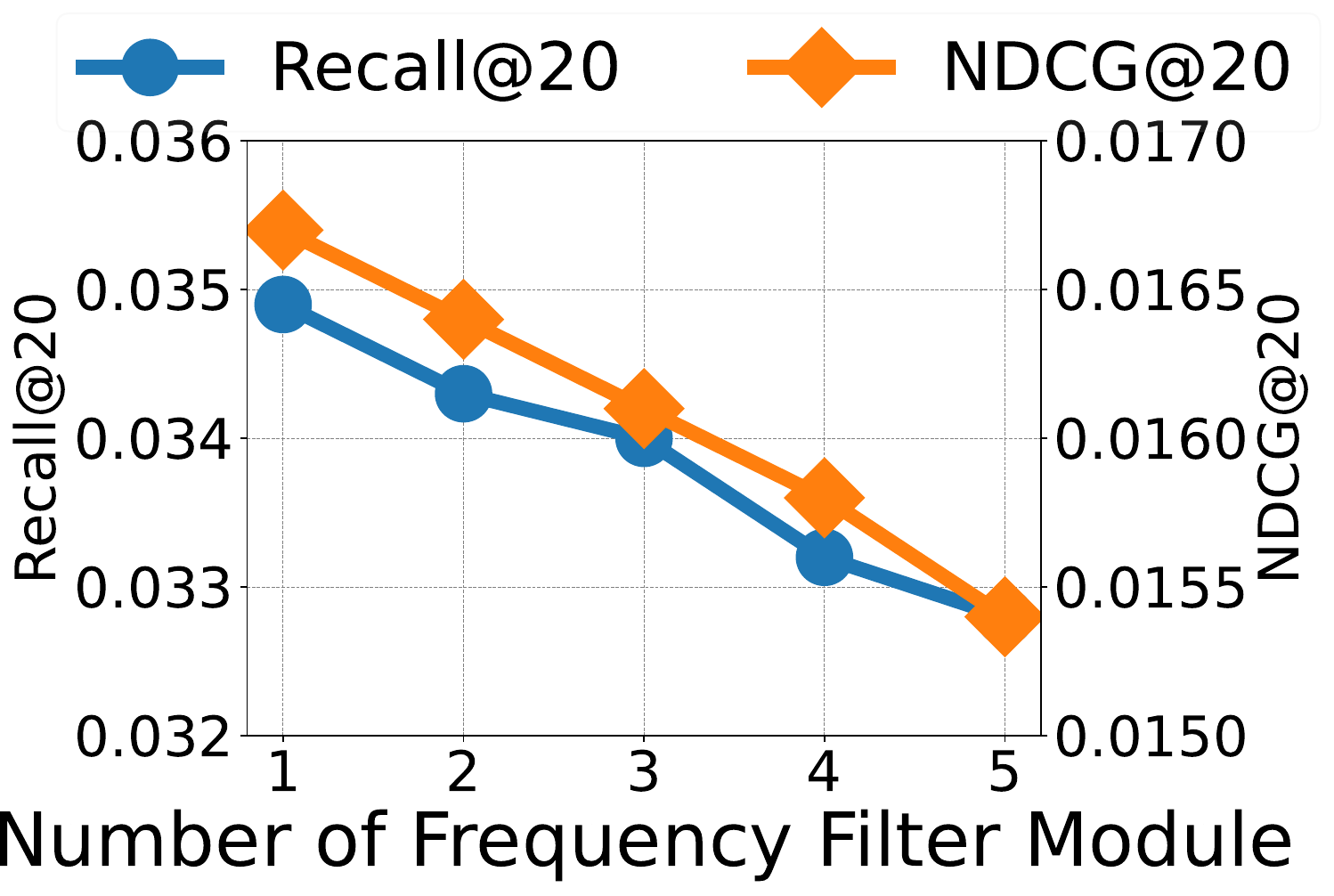}
        \caption{The hyperparameter $N$.}
        \label{fig:layerhyper}
    \end{subfigure}
    \hfill
    \begin{subfigure}{0.22\textwidth}
        \centering
        \includegraphics[width=\linewidth]{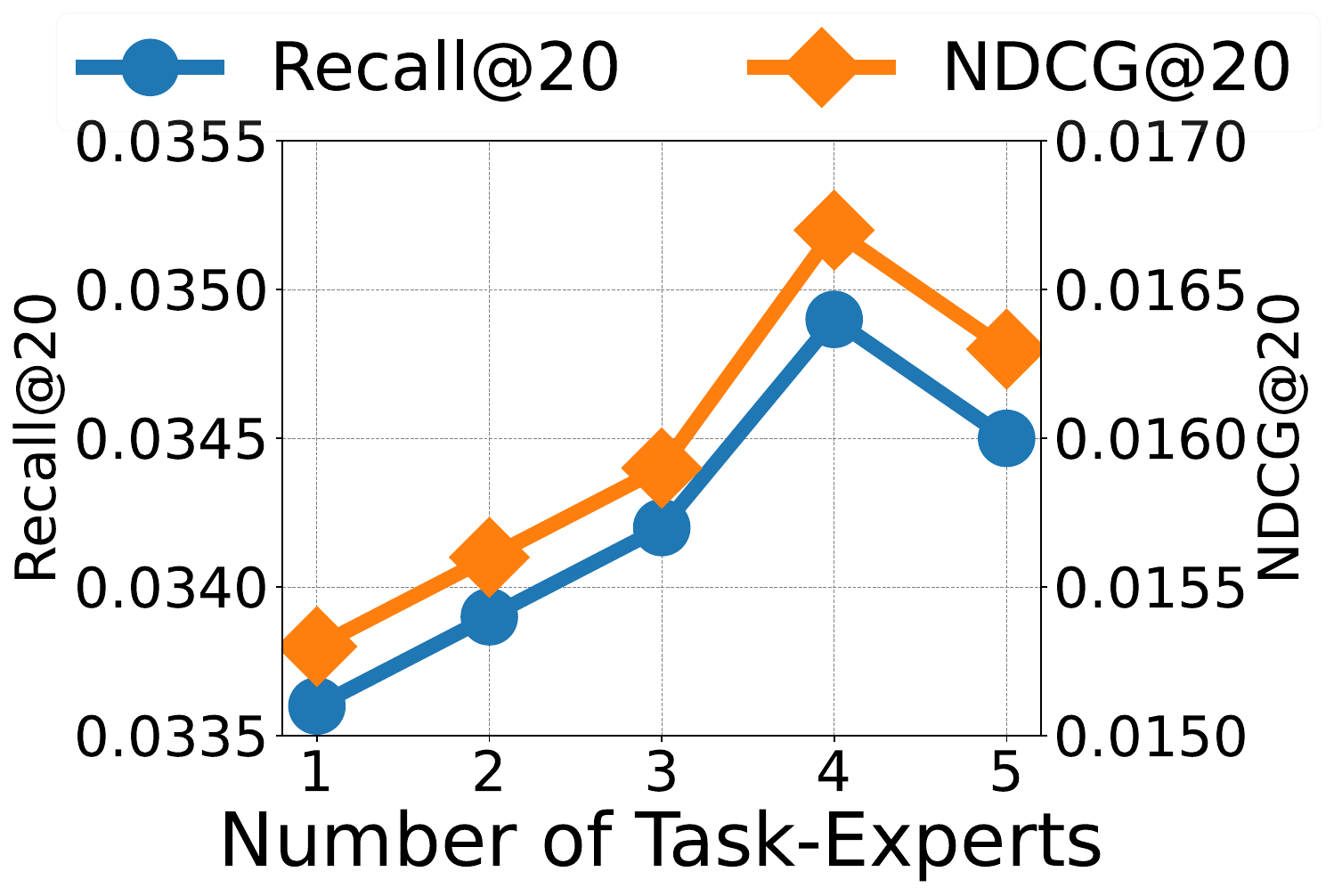}
        \caption{The hyperparameter $E_{m}$.}
        \label{fig:taskhyper}
    \end{subfigure}

    \begin{subfigure}{0.22\textwidth}
        \centering
        \includegraphics[width=\linewidth]{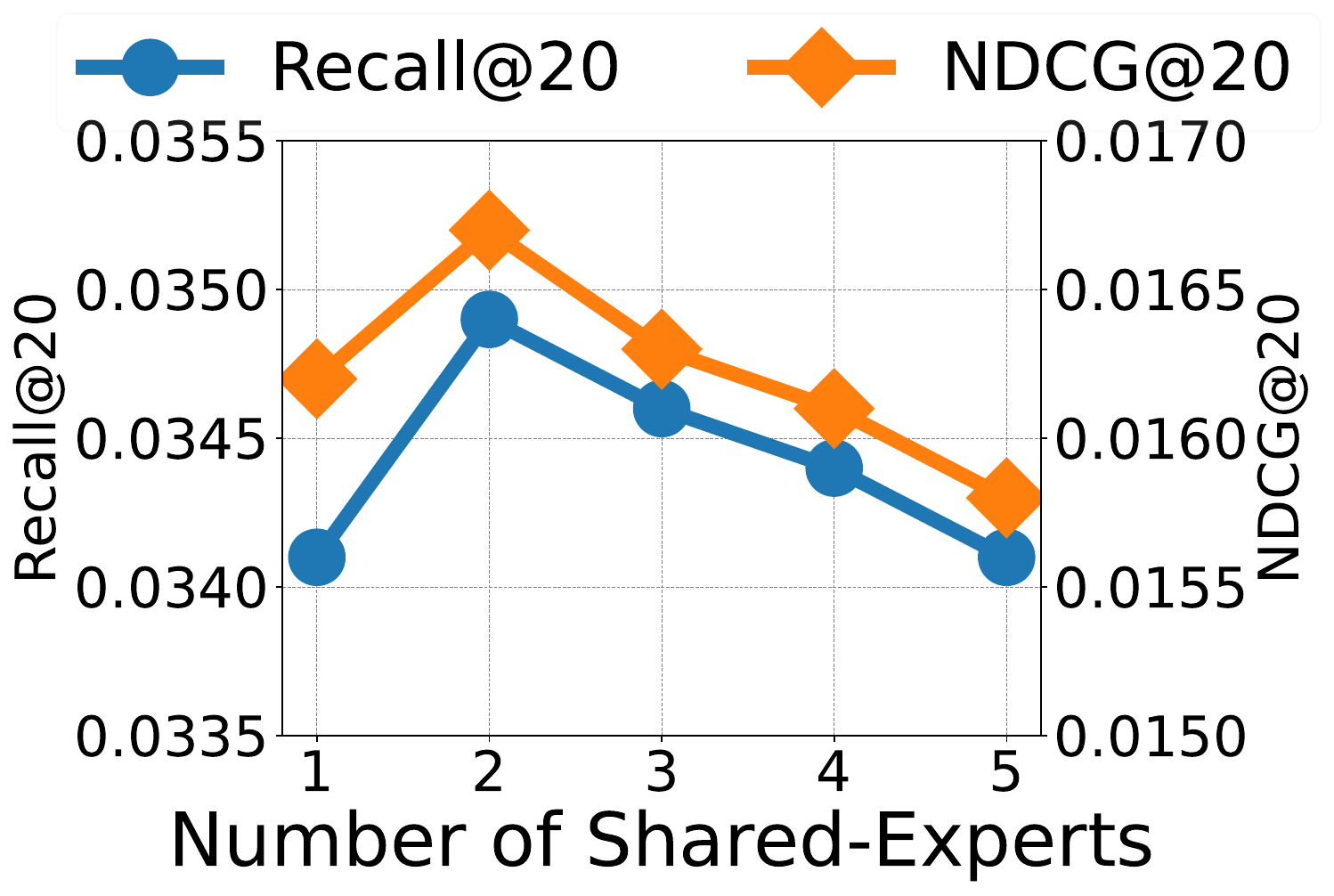}
        \caption{The hyperparameter $E_{s}$.}
        \label{fig:sharehyper}
    \end{subfigure}
    \hfill
    \begin{subfigure}{0.22\textwidth}
        \centering
        \includegraphics[width=\linewidth]{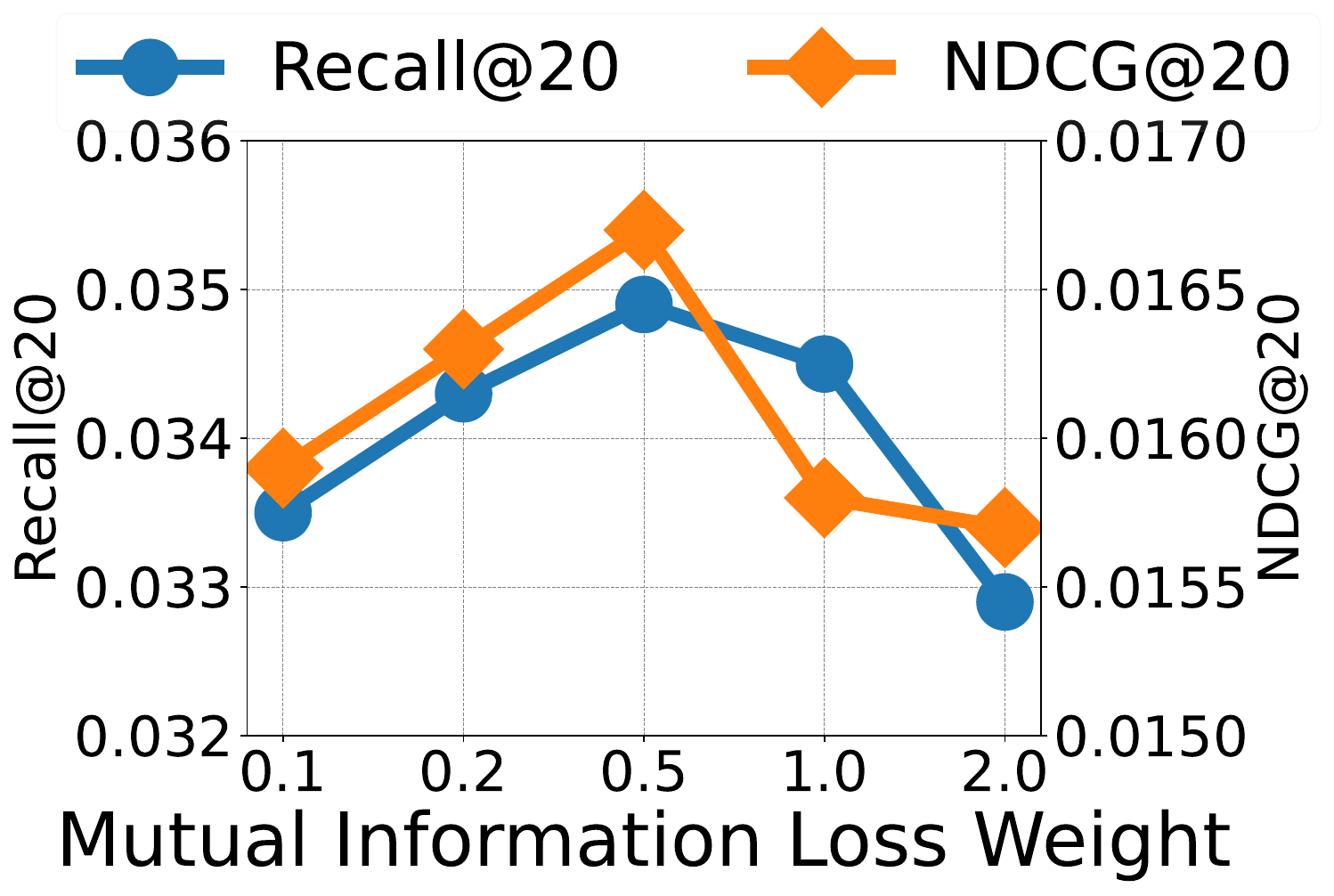}
        \caption{The hyperparameter $\alpha$.}
        \label{fig:alphahyper}
    \end{subfigure}
    \vspace{-0.2cm}
    \caption{Comparison of Hyperparameter Configurations Across Different Metrics in Home Datasets.}
    \Description{The hyperparameter.}
\end{figure}

\textbf{Number of Frequency Filter Modules $N$.}  
As shown in Figure~\ref{fig:layerhyper}, performance consistently declines as $N$ increases from 1 to 5, indicating that one Frequency Filter Module is sufficient to remove irrelevant noise from multimodal representations extracted by pre-trained encoders. Additional filter modules instead introduce over-filtering and weaken signals relevant to user-specific preferences. Therefore, the optimal setting is $N=1$, balancing noise reduction and user-specific preference modeling.

\textbf{Task-Experts ${E}_{m}$.}  
Task-Experts capture modality-specific patterns such as visual textures or textual semantics. Figure~\ref{fig:taskhyper} shows that increasing ${E}_{m}$ from 1 to 4 improves performance by enhancing representational diversity. When ${E}_{m}>4$, performance declines due to redundancy and overlap. The best result is achieved at ${E}_{m}=4$.

\textbf{Share-Experts ${E}_{s}$.}  
Share-Experts integrate common patterns across modalities. As shown in Figure~\ref{fig:sharehyper}, increasing ${E}_{s}$ improves performance up to 2, beyond which overfitting and reduced generalization occur. The setting ${E}_{s}=2$ offers a strong balance between shared information and modality-specific focus.

\textbf{Mutual Information Loss Weight $\alpha$.}  
The hyperparameter $\alpha$ controls the strength of mutual information regularization for routing. Figure~\ref{fig:alphahyper} shows that increasing $\alpha$ from 0.1 to 0.5 consistently improves Recall@20 and NDCG@20. Performance peaks at $\alpha=0.5$, while larger values reduce flexibility. Thus, $\alpha=0.5$ provides the most effective trade-off.

\subsection{Time Complexity (RQ4)}

We analyze the time complexity of SG-UMP from both theoretical and empirical perspectives, with details provided in \textbf{Appendix A.4}. Overall, the complexity is dominated by the sequence modeling component, leading to an approximate complexity of $O(L^2 \cdot d)$, which is consistent with standard self-attention based sequential models. Empirically, incorporating SG-UMP introduces only moderate computational overhead across all backbones. For example, on the Yelp dataset, the epoch time increases by 5.1\% for SASRec and 17.1\% for Oracle4Rec. These results indicate that SG-UMP achieves a trade-off between performance and efficiency, while remaining practical for real-world applications.

%% file: 6-Related_Work.tex
\section{Related Work}
\textbf{Sequential Recommendation.} 
SR leverages the temporal order of user interactions to predict subsequent items. Early methods relied on Markov Chains (MC)~\cite{markov, markov1_FPMC}, such as FPMC~\cite{markov1_FPMC}, which combined MC transitions with matrix factorization, and Fossil~\cite{markov}, which extended to higher-order MCs. With deep learning, models like GRU4Rec~\cite{GRU4Rec}, RNNs~\cite{RNN1, RNN3}, and LSTMs~\cite{LSTM} improved sequential encoding. Transformer-based models~\cite{Transformer, Transformer1}, such as SASRec~\cite{SASRec} and BERT4Rec~\cite{Bert4rec}, utilized self-attention to capture global sequence dependencies, outperforming MC- and RNN-based methods. Recent advancements include STOSA~\cite{STOSA}, which employs Wasserstein Self-Attention for robustness, FMLP~\cite{FMLP} with all-MLP sequence modeling, and Oracle4Rec~\cite{Oracle4Rec} train model with future user interactions to improve the accuracy of SR.

\textbf{Multimodal Recommendation.}
Multimodal recommendation systems leverage heterogeneous data such as images and text to better capture user preferences~\cite{multimodal}. Early studies, such as VBPR~\cite{VBPR} and DeepStyle~\cite{Deepstyle}, incorporate visual features into collaborative filtering frameworks using convolutional networks, while VECF~\cite{VECF} further introduces attention mechanisms to enhance feature interaction. Recent works extend multimodal modeling to SR scenarios. For example, MISSRec~\cite{MISSRec} addresses cold-start issues through pre-training and MMSR~\cite{MMSR} utilizes graph-based structures with adaptive gating mechanisms. Other approaches, such as MMMLP~\cite{MMMLP} and MP4SR~\cite{MP4SR}, explore MLP-based architectures and contrastive learning strategies to improve multimodal representation and fusion. Despite these advances, most existing methods assume uniform user attention across modalities and consistent modality importance across datasets, limiting their ability to capture user-specific preference heterogeneity and adapt to varying modality characteristics. To address this limitation, we propose SG-UMP, a unified plug-and-play multimodal recommendation framework that enables adaptive modeling through flexible module composition and dynamic processing strategies.

%% file: 7-Conclusion.tex
\section{Conclusion}
This paper proposes SG-UMP, a plug-and-play plugin for enhancing multimodal information processing in MSR. To address user-level preference heterogeneity and dataset-level modality bias, SG-UMP introduces a Module Combiner for flexible multimodal processing and a Module Router for dynamic module ordering. Experiments on multiple real-world datasets demonstrate that SG-UMP consistently improves recommendation performance across different backbones and multimodal settings. In future work, we will explore more efficient routing strategies and extend SG-UMP to additional modalities and recommendation scenarios.

%% file: 9-Acknowledge.tex
\section{Acknowledgments}
This work was supported by the National Major Science andTechnology Program of China (No. 2026ZD1307700), Basic Research Program of Jiangsu (Grant No. BK20250668), Jiangsu Provincial Young Science and Technology Talent Support Program (Grants No. JSTJ-2025-944), Science and Technology Major Special Program of Jiangsu (Grants No. BG2024028), Basic Science (Natural Science) Research Project of Jiangsu Province Higher Education Institutions (Grants No. 25KJB-520039).

%% file: 8-Appendix.tex
\appendix
\section{Appendix}

\subsection{Low-Pass Filtering Model}\label{appendix:lowpass}

The Low-Pass Filtering Model refines input sequences by applying Fourier Transform techniques to extract stable low-frequency components while suppressing high-frequency noise. This addresses noisy multimodal sequences (Principle 1) and ensures compatibility through standardized input-output design (Principle 2). The model consists of three stages.

\textbf{Fourier Transform:}  
The input sequence $x_n$ is transformed into the frequency domain using the Fast Fourier Transform (FFT) to decompose it into frequency components, enabling the separation of stable low-frequency patterns from transient high-frequency noise. Formally, $\mathbf{X}_n, \mathbf{f}_n \leftarrow \text{FFT}(x_n)$, where $\mathbf{X}_n \in \mathbb{C}^{c \times d}$ represents the spectrum matrix and $\mathbf{f}_n \in \mathbb{R}^c$ is the frequency distribution vector. The dimension $c$ denotes the number of discrete frequency bins determined by the sequence length or FFT configuration, controlling the frequency resolution. To retain only low-frequency components, a dynamic cutoff frequency $f_{cut}$ is determined based on the $q$-quantile of $\mathbf{f}_n$. An indicator matrix $\mathbf{M}_n = \left[\mathds{1}_{\mathbf{f}_{n,1} < f_{cut}}, \cdots, \mathds{1}_{\mathbf{f}_{n,c} < f_{cut}}\right]^\top \otimes \mathbf{1}^\top$ filters out high-frequency noise, where $\mathds{1}$ selects frequencies below $f_{cut}$ and $\otimes$ denotes the outer product. The refined spectrum matrix is obtained as $\tilde{\mathbf{X}}_n = \mathbf{X}_n \odot \mathbf{M}_n$, where $\odot$ indicates element-wise multiplication to retain only low-frequency components.

\textbf{Inverse Fourier Transform:}  
The filtered spectrum $\tilde{\mathbf{X}}_n$ is transformed back into the time domain using the Inverse Fast Fourier Transform (IFFT) to reconstruct the filtered signal while preserving stable features, expressed as $\tilde{x}_n = \text{IFFT}(\tilde{\mathbf{X}}_n)$.

\textbf{Residual Connection and Layer Normalization:}  
To preserve any valuable high-frequency information, the reconstructed signal $\tilde{x}_n$ is combined with the original input $x_n$ using a residual connection and Layer Normalization, given by $x_{n+1} = \text{LayerNorm}(\tilde{x}_n + x_n)$.

The Low-Pass Filtering Model thus refines input sequences by isolating patterns, suppressing noise, and aligning with user attention for integration with other recommendation modules.

\subsection{Theoretical Analysis of Conditional Mutual Information Routing}
\label{appendix:theory_router}

\subsubsection{Preliminary on Partial Information Decomposition Theory}
Classical information theory quantifies how much one variable informs another, but its direct extension to multiple sources remains challenging. To address this limitation, the \textbf{Partial Information Decomposition (PID)} framework~\cite{wollstadt2023, liang2023, liang2024, Bertschinger2014} decomposes the total information that two sources $(X_1, X_2)$ provide about a target $Y$ into four distinct components: 

\begin{equation}
\begin{aligned}
I(X_1, X_2; Y)
&= I_{\text{unique}}(X_1; Y)
+ I_{\text{unique}}(X_2; Y) \\
&\quad + I_{\text{redundant}}(X_1, X_2; Y)
+ I_{\text{synergy}}(X_1, X_2; Y).
\end{aligned}
\end{equation}

\textbf{Unique information} refers to the knowledge about $Y$ that is provided only by $X_1$ or $X_2$.  
\textbf{Redundant information} represents the overlap between $X_1$ and $X_2$ about $Y$, while  
\textbf{Synergy information} denotes the information that emerges only when both $X_1$ and $X_2$ are jointly observed.

Formally, synergy can be defined as the difference between the total mutual information under the true distribution $p$ and the minimum achievable mutual information under all distributions $q \in \Delta_p$ that preserve the pairwise marginals between each modality and the target~\cite{Bertschinger2014, liang2023}:

\begin{align}
S &= I_p(X_1, X_2; Y) - \min_{q \in \Delta_p} I_q(X_1, X_2; Y),\\
\Delta_p &= \{ q \in \Delta : q(x_i, y) = p(x_i, y),\; i \in \{1, 2\} \}.
\end{align}

Intuitively, $S$ measures the gain of information that arises when both modalities interact, beyond what each modality provides independently. In the context of multimodal learning, this quantity represents the \textit{collaborative effect between modalities}, indicating how much additional information emerges through their joint consideration.

\subsubsection{Connection to Conditional Mutual Information in the Router}
Building on PID theory, the \textbf{Conditional Mutual Information (CMI)} serves as an operational form of synergy measurement under specific conditioning variables. For two random variables $M$ (module) and $m$ (modality), conditioned on the input embedding $\mathbf{e}_i$, CMI is defined as:

\begin{equation}
\begin{aligned}
I(M; m \mid \mathbf{e}_i)
&= \sum_{k,m} 
P(M_k, m \mid \mathbf{e}_i)
\log 
\frac{P(M_k, m \mid \mathbf{e}_i)}
{P(M_k \mid \mathbf{e}_i) P(m \mid \mathbf{e}_i)}.
\end{aligned}
\end{equation}

This term measures how much knowing the modality $m$ reduces uncertainty about the module assignment $M$, given input $\mathbf{e}_i$. In our framework, maximizing $I(M; m \mid \mathbf{e}_i)$ encourages the router to discover and exploit modality–module dependencies, thereby reflecting the \textit{synergistic interactions between modalities and functional modules}. More concretely:

\begin{itemize}
    \item When $I(M; m \mid \mathbf{e}_i)$ is high, routing decisions depend strong-ly on modality-specific cues, indicating successful synergy capture.
    \item When $I(M; m \mid \mathbf{e}_i)$ is low, routing becomes modality agnostic, leading to uniform and degenerate assignments.
\end{itemize}

Therefore, maximizing CMI emphasizes the synergistic component of information flow between modalities and modules, establishing a direct theoretical bridge between PID-based information decomposition and our information-aware router design. In summary, PID provides the theoretical foundation for analyzing information interactions across modalities, while CMI translates this concept into a learnable, conditional form suitable for routing optimization. The proposed mutual information regularization thus ensures that each module captures distinct, modality-aligned, and synergistic information, preventing redundant or degenerate routing behaviors.

\subsection{Algorithm}\label{alg:sgump}

\begin{algorithm}[!ht]
\caption{Training Procedure of SG-UMP Framework}
\begin{algorithmic}[1]
\Require Training data $\mathcal{D} = \{(u, i^+, i^-)\}$; multimodal embeddings $\mathbf{e}^{id}, \mathbf{e}^{img}, \mathbf{e}^{txt}$; modules $\{M_1, M_2, M_3\}$; router parameters $\Theta_r$; hyperparameter $\alpha$
\Ensure Updated router and module parameters
\For{each mini-batch $(u, i^+, i^-)$}
    \State \textbf{Feature Encoding:} 
    Compute modality-specific representations $\mathbf{h}^m = f^m(\mathbf{e}^m)$ for $m \in \{id, img, txt\}$
    \State \textbf{Dispatch Score:} 
    For each module $M_k$, compute $\mathbf{s}_k^m = (\mathbf{w}_k)^{\top}\mathrm{ReLU}(\mathbf{W}^{m}\mathbf{h}^m + \mathbf{b}_k)$
    \State Aggregate scores $\mathbf{s} = \sum_m \mathbf{s}^m$ and normalize $\mathbf{D} = \mathrm{softmax}(\mathbf{s})$
    \State \textbf{Routing:} Sort $\mathbf{s}$ in descending order $\rightarrow$ routing order $\pi = (\pi_1, \pi_2, \pi_3)$
    \State Concatenate multimodal embeddings $\mathbf{z} = [\mathbf{h}^{id}; \mathbf{h}^{img}; \mathbf{h}^{txt}]$
    \State Sequentially apply modules $\mathbf{y} = M_{\pi_3}(M_{\pi_2}(M_{\pi_1}(\mathbf{z})))$
    \State \textbf{Recommendation Loss:} 
    \State Compute $\mathcal{L}_{\text{rec}} = -\log \sigma(r(u, i^+) - r(u, i^-))$
    \State \textbf{Mutual Information Loss:} 
    \State Estimate $P(M_k, m \mid \mathbf{e})$, $P(M_k \mid \mathbf{e})$, and $P(m \mid \mathbf{e})$
    \State Compute $\mathcal{L}_{MI} = - \sum_{k,m} P(M_k, m \mid \mathbf{e}) \log \frac{P(M_k, m \mid \mathbf{e})}{P(M_k \mid \mathbf{e}) P(m \mid \mathbf{e})}$
    \State \textbf{Joint Optimization:} 
    $\mathcal{L} = \mathcal{L}_{\text{rec}} + \alpha \mathcal{L}_{MI}$
\EndFor
\end{algorithmic}
\end{algorithm}

\subsection{Time Complexity (RQ4)}\label{time_complexity}

\begin{table*}[ht!]
\centering
\caption{
Average training time per epoch (in seconds) for baseline models and their SG-UMP enhanced variants. Here, \textbf{w/o SG-UMP} denotes the baseline model without our proposed module, while \textbf{w/ SG-UMP} indicates the model integrated with SG-UMP. \textbf{Bold values} represent the training time of the SG-UMP variants. All results are averaged over five runs to ensure robustness and are statistically significant with \(p < 0.05\).
}
\begin{tabular}{c|cc|cc|cc}
\toprule
\multirow{2}{*}{\textbf{Dataset}} 
& \multicolumn{2}{c|}{\textbf{SASRec}} 
& \multicolumn{2}{c|}{\textbf{STOSA}} 
& \multicolumn{2}{c}{\textbf{Oracle4Rec}} \\
\cmidrule(lr){2-7}
& \textbf{w/o SG-UMP} & \textbf{w/ SG-UMP} 
& \textbf{w/o SG-UMP} & \textbf{w/ SG-UMP} 
& \textbf{w/o SG-UMP} & \textbf{w/ SG-UMP} \\
\midrule
Home   & 20.35s & \textbf{29.23s} & 24.99s & \textbf{34.88s} & 71.41s & \textbf{84.88s} \\
Beauty & 5.75s  & \textbf{6.39s}  & 7.40s  & \textbf{8.22s}  & 24.29s & \textbf{28.22s} \\
Office & 0.48s  & \textbf{0.72s}  & 0.63s  & \textbf{1.13s}  & 7.35s  & \textbf{15.11s} \\
Yelp   & 735.32s & \textbf{772.53s} & 791.03s & \textbf{828.08s} & 1242.51s & \textbf{1455.08s} \\
\bottomrule
\end{tabular}
\label{tab:time}
\end{table*}

We analyze the time complexity of SG-UMP from both theoretical and empirical perspectives. The theoretical complexity is determined by its four main components. Let $L$ denote the sequence length, $d$ the embedding dimension, $K$ the number of modules (i.e., $K=3$), and $N_m$ the number of modalities (i.e., $N_m=3$). (1) \textit{Frequency Filter Layer:} Applies FFT and IFFT to multimodal embeddings. Each layer has complexity $O(L \cdot d \log L)$, and with $N$ such layers, the total is $O(N \cdot L \cdot d \log L)$~\cite{FFT1, FFT2}. (2) \textit{Hierarchical Attention Layer:} Each attention block has complexity $O(L^2 \cdot d)$, resulting in $O(M \cdot L^2 \cdot d)$ with $M$ blocks. (3) \textit{Multi-Scale Fusion Layer:} Each sequence element passes through $K$ experts and a gating network, yielding $O(L \cdot K \cdot d)$ per layer. With $P$ layers, the cost becomes $O(P \cdot L \cdot K \cdot d)$. (4) \textit{Module Router:} Modality-aware gating and score sorting require $O(d^2)$ due to small $K$ and $N_m$ constants. Summing all components, the total complexity is \(
O(N \cdot L \cdot d \log L + M \cdot L^2 \cdot d + P \cdot L \cdot K \cdot d + d^2).\)
In practice, especially when $L \gg d$, the attention term $M \cdot L^2 \cdot d$ dominates. Even when $d > L$, the router's cost $O(d^2)$ remains minor relative to sequence-level modules. Thus, the overall complexity is approximately $O(L^2 \cdot d)$. To validate this, Table~\ref{tab:time} reports the average epoch time on four datasets. Incorporating SG-UMP introduces moderate computational overhead across all backbones. For example, on the large-scale Yelp dataset, epoch time increases by only 5.1\% for SASRec and 17.1\% for Oracle4Rec. Similar trends are observed for smaller datasets, with consistent increases proportional to model complexity. These results confirm the practicality of SG-UMP. The performance gains brought by richer multimodal understanding justify the additional cost, and the runtime remains feasible for real-world applications. SG-UMP offers a favorable trade-off between accuracy and efficiency.

\subsection{Baselines}

\textbf{Baselines:} We compare our proposed model with recent representative recommendation systems, categorized into two groups:

\textbf{(1) Traditional Recommendation:}  

\begin{itemize}
    \item SASRec\footnote{\url{https://github.com/kang205/SASRec}}~\cite{SASRec}: Uses the Transformer architecture for sequential modeling.
    \item BERT4-Rec\footnote{\url{https://github.com/FeiSun/BERT4Rec}}~\cite{Bert4rec}: Employs a bidirectional Transformer with a objective.
    \item LightGCN\footnote{\url{https://github.com/enoche/MMRec}}~\cite{LightGCN}: Simplifies GCNs by focusing on neighborhood aggregation for recommendation.
    \item STOSA\footnote{\url{https://github.com/zfan20/STOSA}}~\cite{STOSA}: Models items as Gaussian distributions to encode uncertainty.
    \item FMLP-Rec\footnote{\url{https://github.com/Woeee/FMLP-Rec}}~\cite{FMLP}: Uses frequency-domain filters in an all-MLP architecture.
    \item Oracle4Rec\footnote{\url{https://github.com/Yaveng/Oracle4Rec}}~\cite{Oracle4Rec}: Guides model training with future user interactions to improve the accuracy of sequential recommendation.
\end{itemize}
\textbf{(2) Multimodal Recommendation:}  
\begin{itemize}
    \item VBPR\footnote{\url{https://github.com/enoche/MMRec}}~\cite{VBPR}: Extends MF-BPR~\cite{MFBPR} by incorporating visual features.
    \item MMMLP\footnote{\url{https://github.com/Applied-Machine-Learning-Lab/MMMLP}}~\cite{MMMLP}: Combines Feature Mixer, Fusion Mixer, and Prediction layers for efficient multimodal learning.
    \item MML\footnote{\url{https://github.com/RUCAIBox/MML}}~\cite{MML}: Uses multimodal meta-learners with an adaptive fusion layer.
    \item MMSR\footnote{\url{https://github.com/HoldenHu/MMSR}}~\cite{MMSR}: Balances intra- and inter-modality fusion using a graph-based approach.
    \item MMSBR\footnote{\url{https://github.com/Zhang-xiaokun/MMSBR}}~\cite{MMSBR}: Applies contrastive learning to reduce multimodal noise.
    \item MP4SR\footnote{\url{https://github.com/lzz0007/MP4SR}}~\cite{MP4SR}: Encodes multimodal sequences by transforming item images into text.
\end{itemize}

%% file: MM26.bib
@inproceedings{liang2023,
    author = {Liang, Paul Pu and Cheng, Yun and Fan, Xiang and others},
    title = {Quantifying \& modeling multimodal interactions: an information decomposition framework},
    year = {2023},
    publisher = {Curran Associates Inc.},
    address = {Red Hook, NY, USA},
    booktitle = {Proceedings of the 37th International Conference on Neural Information Processing Systems},
    numpages = {43},
    series = {NIPS '23}
}

@article{liang2024,
    author = {Liang, Paul Pu and Zadeh, Amir and Morency, Louis-Philippe},
    title = {Foundations \& Trends in Multimodal Machine Learning: Principles, Challenges, and Open Questions},
    year = {2024},
    publisher = {Association for Computing Machinery},
    address = {New York, NY, USA},
    volume = {56},
    number = {10},
    journal = {ACM Comput. Surv.},
    numpages = {42},
}

@article{wollstadt2023,
    author = {Wollstadt, Patricia and Schmitt, Sebastian and Wibral, Michael},
    title = {A rigorous information-theoretic definition of redundancy and relevancy in feature selection based on (partial) information decomposition},
    year = {2023},
    publisher = {JMLR.org},
    volume = {24},
    number = {1},
    journal = {J. Mach. Learn. Res.},
    numpages = {44},
}

@article{bertschinger2014,
  title={Quantifying unique information},
  author={Bertschinger, Nils and Rauh, Johannes and Olbrich, Eckehard and Jost, J{\"u}rgen and Ay, Nihat},
  journal={Entropy},
  volume={16},
  number={4},
  pages={2161--2183},
  year={2014},
  publisher={Multidisciplinary Digital Publishing Institute}
}

@inproceedings{MDSBR,
    author = {Li, Yutong and Zhang, Xinyi},
    title = {MDSBR: Multimodal Denoising for Session-based Recommendation},
    year = {2025},
    publisher = {Association for Computing Machinery},
    address = {New York, NY, USA},
    booktitle = {Proceedings of the Nineteenth ACM Conference on Recommender Systems},
    pages = {268–278},
    numpages = {11},
    series = {RecSys '25}
}

@inproceedings{BeFA,
author = {Fan, Qile and Yu, Penghang and Tan, Zhiyi and Bao, Bing-Kun and Lu, Guanming},
title = {BeFA: a general behavior-driven feature adapter for multimedia recommendation},
year = {2025},
publisher = {AAAI Press},
booktitle = {Proceedings of the Thirty-Ninth AAAI Conference on Artificial Intelligence},
articleno = {1293},
numpages = {11},
address = {Washington, USA},
series = {AAAI'25}
}

@article{LSTM,
  title={Long short-term memory},
  author={Hochreiter, Sepp and Schmidhuber, J{\"u}rgen},
  journal={Neural computation},
  volume={9},
  number={8},
  pages={1735--1780},
  year={1997},
  publisher={MIT press}
}

@inproceedings{GRU4Rec,
  author       = {Bal{\'{a}}zs Hidasi and
                  Alexandros Karatzoglou and
                  Linas Baltrunas and
                  Domonkos Tikk},
  editor       = {Yoshua Bengio and
                  Yann LeCun},
  title        = {Session-based Recommendations with Recurrent Neural Networks},
  booktitle    = {4th International Conference on Learning Representations, {ICLR} 2016,
                  San Juan, Puerto Rico, May 2-4, 2016, Conference Track Proceedings},
  year         = {2016},
  publisher = {ICLR},            
  address   = {San Juan, Puerto Rico},  
  pages    = {} 
}

@inproceedings{Transformer,
author = {Vaswani, Ashish and Shazeer, Noam and Parmar, Niki and Uszkoreit, Jakob and Jones, Llion and Gomez, Aidan N. and Kaiser, \L{}ukasz and Polosukhin, Illia},
title = {Attention is all you need},
year = {2017},
publisher = {Curran Associates Inc.},
address = {Red Hook, NY, USA},
booktitle = {Proceedings of the 31st International Conference on Neural Information Processing Systems},
pages = {6000–6010},
numpages = {11},
series = {NIPS'17}
}

@inproceedings{Bert,
  author       = {Jacob Devlin and
                  Ming{-}Wei Chang and
                  Kenton Lee and
                  Kristina Toutanova},
  title        = {{BERT:} Pre-training of Deep Bidirectional Transformers for Language
                  Understanding},
  booktitle    = {Proceedings of the 2019 Conference of the North American Chapter of
                  the Association for Computational Linguistics: Human Language Technologies,
                  {NAACL-HLT} 2019, Minneapolis, MN, USA, June 2--7, 2019, Volume 1 (Long
                  and Short Papers)},
  pages        = {4171--4186},
  publisher    = {Association for Computational Linguistics},
  address      = {Minneapolis, Minnesota, USA},
  year         = {2019},
}

@inproceedings{SASRec,
  author       = {Wang{-}Cheng Kang and
                  Julian J. McAuley},
  title        = {Self-Attentive Sequential Recommendation},
  booktitle    = {{IEEE} International Conference on Data Mining, {ICDM} 2018, Singapore,
                  November 17-20, 2018},
  pages        = {197--206},
  publisher    = {{IEEE} Computer Society},
  year         = {2018},
  address     = {Singapore},
}

@inproceedings{Bert4rec,
author = {Sun, Fei and Liu, Jun and Wu, Jian and Pei, Changhua and Lin, Xiao and Ou, Wenwu and Jiang, Peng},
title = {BERT4Rec: Sequential Recommendation with Bidirectional Encoder Representations from Transformer},
year = {2019},
publisher = {Association for Computing Machinery},
address = {New York, NY, USA},
booktitle = {Proceedings of the 28th ACM International Conference on Information and Knowledge Management},
pages = {1441–1450},
numpages = {10},
series = {CIKM '19}
}

@inproceedings{markov,
  author    = {He, Ruining and McAuley, Julian},
  title     = {Fusing similarity models with Markov chains for sparse sequential recommendation},
  booktitle = {2016 IEEE 16th International Conference on Data Mining (ICDM)},
  year      = {2016},
  pages     = {191--200},
  address   = {Barcelona, Spain},
  publisher    = {{IEEE} Computer Society},
}

@inproceedings{markov1_FPMC,
    author = {Rendle, Steffen and Freudenthaler, Christoph and Schmidt-Thieme, Lars},
    title = {Factorizing personalized Markov chains for next-basket recommendation},
    year = {2010},
    publisher = {Association for Computing Machinery},
    address = {New York, NY, USA},
    booktitle = {Proceedings of the 19th International Conference on World Wide Web},
    pages = {811–820},
    numpages = {10},
    series = {WWW '10}
}

@inproceedings{Transformer1,
    title = "{PDALN}: Progressive Domain Adaptation over a Pre-trained Model for Low-Resource Cross-Domain Named Entity Recognition",
    author = "Zhang, Tao  and
      Xia, Congying  and
      Yu, Philip S.  and
      Liu, Zhiwei  and
      Zhao, Shu",
    booktitle = "Proceedings of the 2021 Conference on Empirical Methods in Natural Language Processing",
    month = nov,
    year = "2021",
    address = "Online and Punta Cana, Dominican Republic",
    publisher = "Association for Computational Linguistics",
    pages = "5441--5451",
}

@inproceedings{RNN1,
  title        = {Hierarchical Gating Networks for Sequential Recommendation},
  author       = {Ma, Chen and Kang, Peng and Liu, Xue},
  booktitle    = {Proceedings of the 25th ACM SIGKDD International Conference on Knowledge Discovery and Data Mining (KDD '19)},
  pages        = {825--833},
  address      = {New York, NY, USA},
  publisher    = {Association for Computing Machinery},
  year         = {2019},
}

@inproceedings{RNN3,
  title        = {Personalizing Session-based Recommendations with Hierarchical Recurrent Neural Networks},
  author       = {Quadrana, Massimo and Karatzoglou, Alexandros and Hidasi, Bal{\'a}zs and Cremonesi, Paolo},
  booktitle    = {Proceedings of the Eleventh ACM Conference on Recommender Systems (RecSys '17)},
  pages        = {130--137},
  publisher    = {Association for Computing Machinery},
  address      = {New York, NY, USA},
  year         = {2017},
}

@article{multimodal,
  title={Multimodal machine learning: A survey and taxonomy},
  author={Baltru{\v{s}}aitis, Tadas and Ahuja, Chaitanya and Morency, Louis-Philippe},
  journal={IEEE transactions on pattern analysis and machine intelligence},
  volume={41},
  number={2},
  pages={423--443},
  year={2018},
  publisher={IEEE}
}

@inproceedings{VBPR,
author = {He, Ruining and McAuley, Julian},
title = {VBPR: visual Bayesian Personalized Ranking from implicit feedback},
year = {2016},
publisher = {AAAI Press},
booktitle = {Proceedings of the Thirtieth AAAI Conference on Artificial Intelligence},
pages = {144–150},
numpages = {7},
address = {Phoenix, Arizona},
series = {AAAI'16}
}

@inproceedings{Deepstyle,
author = {Liu, Qiang and Wu, Shu and Wang, Liang},
title = {DeepStyle: Learning User Preferences for Visual Recommendation},
year = {2017},
publisher = {Association for Computing Machinery},
address = {New York, NY, USA},
booktitle = {Proceedings of the 40th International ACM SIGIR Conference on Research and Development in Information Retrieval},
pages = {841–844},
numpages = {4},
series = {SIGIR '17}
}

@inproceedings{VECF,
author = {Chen, Xu and Chen, Hanxiong and Xu, Hongteng and Zhang, Yongfeng and Cao, Yixin and Qin, Zheng and Zha, Hongyuan},
title = {Personalized Fashion Recommendation with Visual Explanations based on Multimodal Attention Network: Towards Visually Explainable Recommendation},
year = {2019},
publisher = {Association for Computing Machinery},
address = {New York, NY, USA},
booktitle = {Proceedings of the 42nd International ACM SIGIR Conference on Research and Development in Information Retrieval},
pages = {765–774},
numpages = {10},
series = {SIGIR'19}
}

@inproceedings{STOSA,
    author = {Fan, Ziwei and Liu, Zhiwei and Wang, Yu and Wang, Alice and Nazari, Zahra and Zheng, Lei and Peng, Hao and Yu, Philip S.},
    title = {Sequential Recommendation via Stochastic Self-Attention},
    year = {2022},
    publisher = {Association for Computing Machinery},
    address = {New York, NY, USA},
    booktitle = {Proceedings of the ACM Web Conference 2022},
    pages = {2036–2047},
    numpages = {12},
    series = {WWW '22}
}

@ARTICLE{MMSBR,
  author={Zhang, Xiaokun and Xu, Bo and Ma, Fenglong and Li, Chenliang and Yang, Liang and Lin, Hongfei},
  journal={IEEE Transactions on Knowledge and Data Engineering}, 
  title={Beyond Co-Occurrence: Multi-Modal Session-Based Recommendation}, 
  year={2024},
  volume={36},
  number={4},
  pages={1450-1462},
}

@inproceedings{PLE,
  title        = {Progressive Layered Extraction (PLE): A Novel Multi-Task Learning (MTL) Model for Personalized Recommendations},
  author       = {Tang, Hongyan and Liu, Junning and Zhao, Ming and Gong, Xudong},
  booktitle    = {Proceedings of the 14th ACM Conference on Recommender Systems (RecSys '20)},
  pages        = {269--278},
  publisher    = {Association for Computing Machinery},
  address      = {New York, NY, USA},
  year         = {2020},
}

@inproceedings{FMLP,
  title        = {Filter-Enhanced MLP is All You Need for Sequential Recommendation},
  author       = {Zhou, Kun and Yu, Hui and Zhao, Wayne Xin and Wen, Ji-Rong},
  booktitle    = {Proceedings of the ACM Web Conference 2022 (The Web Conference '22 / WWW '22)},
  pages        = {2388--2399},
  publisher    = {Association for Computing Machinery},
  address      = {New York, NY, USA},
  year         = {2022},
}

@inproceedings{MISSRec,
author = {Wang, Jinpeng and Zeng, Ziyun and Wang, Yunxiao and Wang, Yuting and Lu, Xingyu and Li, Tianxiang and Yuan, Jun and Zhang, Rui and Zheng, Hai-Tao and Xia, Shu-Tao},
title = {MISSRec: Pre-training and Transferring Multi-modal Interest-aware Sequence Representation for Recommendation},
year = {2023},
publisher = {Association for Computing Machinery},
address = {New York, NY, USA},
booktitle = {Proceedings of the 31st ACM International Conference on Multimedia},
pages = {6548–6557},
numpages = {10},
series = {MM '23}
}

@misc{PreTrain_Noise1,
      title={Text Is All You Need: Learning Language Representations for Sequential Recommendation}, 
      author={Jiacheng Li and Ming Wang and Jin Li and Jinmiao Fu and Xin Shen and Jingbo Shang and Julian McAuley},
      year={2023},
      archivePrefix={arXiv},
      primaryClass={cs.IR}
}

@inproceedings{CLIP,
  title={Learning transferable visual models from natural language supervision},
  author={Radford, Alec and Kim, Jong Wook and Hallacy, Chris and Ramesh, Aditya and Goh, Gabriel and Agarwal, Sandhini and Sastry, Girish and Askell, Amanda and Mishkin, Pamela and Clark, Jack and others},
  booktitle={International conference on machine learning},
  pages={8748--8763},
  year={2021},
  publisher={PMLR},
  address = {Virtual}
}

@inproceedings{Yelp1,
author = {Zhang, Xiaokun and Xu, Bo and Wu, Youlin and Zhong, Yuan and Lin, Hongfei and Ma, Fenglong},
title = {FineRec: Exploring Fine-grained Sequential Recommendation},
year = {2024},
publisher = {Association for Computing Machinery},
address = {New York, NY, USA},
booktitle = {Proceedings of the 47th International ACM SIGIR Conference on Research and Development in Information Retrieval},
pages = {1599–1608},
numpages = {10},
series = {SIGIR '24}
}

@inproceedings{Yelp2,
author = {Li, Xuewei and Sun, Aitong and Zhao, Mankun and Yu, Jian and Zhu, Kun and Jin, Di and Yu, Mei and Yu, Ruiguo},
title = {Multi-Intention Oriented Contrastive Learning for Sequential Recommendation},
year = {2023},
publisher = {Association for Computing Machinery},
address = {New York, NY, USA},
booktitle = {Proceedings of the Sixteenth ACM International Conference on Web Search and Data Mining},
pages = {411–419},
numpages = {9},
series = {WSDM '23}
}

@inproceedings{UniSRec,
author = {Hou, Yupeng and Mu, Shanlei and Zhao, Wayne Xin and Li, Yaliang and Ding, Bolin and Wen, Ji-Rong},
title = {Towards Universal Sequence Representation Learning for Recommender Systems},
year = {2022},
publisher = {Association for Computing Machinery},
address = {New York, NY, USA},
booktitle = {Proceedings of the 28th ACM SIGKDD Conference on Knowledge Discovery and Data Mining},
pages = {585–593},
numpages = {9},
series = {KDD '22}
}

@inproceedings{MMSR,
    author = {Hu, Hengchang and Guo, Wei and Liu, Yong and Kan, Min-Yen},
    title = {Adaptive Multi-Modalities Fusion in Sequential Recommendation Systems},
    year = {2023},
    publisher = {Association for Computing Machinery},
    address = {New York, NY, USA},
    booktitle = {Proceedings of the 32nd ACM International Conference on Information and Knowledge Management},
    pages = {843–853},
    numpages = {11},
    series = {CIKM '23}
}

@article{MP4SR,
author = {Zhang, Lingzi and Zhou, Xin and Zeng, Zhiwei and Shen, Zhiqi},
title = {Multimodal Pre-training for Sequential Recommendation via Contrastive Learning},
year = {2024},
publisher = {Association for Computing Machinery},
address = {New York, NY, USA},
volume = {3},
number = {1},
journal = {ACM Trans. Recomm. Syst.},
numpages = {23},
}

@inproceedings{MML,
author = {Pan, Xingyu and Chen, Yushuo and Tian, Changxin and Lin, Zihan and Wang, Jinpeng and Hu, He and Zhao, Wayne Xin},
title = {Multimodal Meta-Learning for Cold-Start Sequential Recommendation},
year = {2022},
publisher = {Association for Computing Machinery},
address = {New York, NY, USA},
booktitle = {Proceedings of the 31st ACM International Conference on Information \& Knowledge Management},
pages = {3421–3430},
numpages = {10},
series = {CIKM '22}
}

@inproceedings{MMMLP,
author = {Liang, Jiahao and Zhao, Xiangyu and Li, Muyang and Zhang, Zijian and Wang, Wanyu and Liu, Haochen and Liu, Zitao},
title={Mmmlp: Multi-modal multilayer perceptron for sequential recommendations},
year = {2023},
publisher = {Association for Computing Machinery},
address = {New York, NY, USA},
booktitle = {Proceedings of the ACM Web Conference 2023},
pages = {1109–1117},
numpages = {9},
series = {WWW '23}
}

@INPROCEEDINGS {CAM1,
author = { Zhou, Bolei and Khosla, Aditya and Lapedriza, Agata and Oliva, Aude and Torralba, Antonio },
booktitle = { 2016 IEEE Conference on Computer Vision and Pattern Recognition (CVPR) },
title = {{ Learning Deep Features for Discriminative Localization }},
year = {2016},
volume = {},
ISSN = {1063-6919},
pages = {2921-2929},
publisher = {IEEE Computer Society},
address = {Los Alamitos, CA, USA},}

@article{CAM2,
author = {Selvaraju, Ramprasaath R. and Cogswell, Michael and Das, Abhishek and Vedantam, Ramakrishna and Parikh, Devi and Batra, Dhruv},
title = {Grad-CAM: Visual Explanations from Deep Networks via Gradient-Based Localization},
year = {2020},
publisher = {Kluwer Academic Publishers},
address = {USA},
volume = {128},
number = {2},
journal = {Int. J. Comput. Vision},
pages = {336–359},
numpages = {24},
}

@INPROCEEDINGS {CAM3,
author = { Wang, Haofan and Wang, Zifan and Du, Mengnan and Yang, Fan and Zhang, Zijian and Ding, Sirui and Mardziel, Piotr and Hu, Xia },
booktitle = { 2020 IEEE/CVF Conference on Computer Vision and Pattern Recognition Workshops (CVPRW) },
title = {{ Score-CAM: Score-Weighted Visual Explanations for Convolutional Neural Networks }},
year = {2020},
volume = {},
pages = {111-119},
publisher = {IEEE Computer Society},
address = {Los Alamitos, CA, USA},}

@article{FFT1,
  title={Fast Fourier transforms: a tutorial review and a state of the art},
  author={Duhamel, Pierre and Vetterli, Martin},
  journal={Signal processing},
  volume={19},
  number={4},
  pages={259--299},
  year={1990},
  publisher={Elsevier}
}

@inproceedings{FFT2,
    author = {Yu, Wenhui and Qin, Zheng},
    title = {Graph convolutional network for recommendation with low-pass collaborative filters},
    year = {2020},
    publisher = {JMLR.org},
    booktitle = {Proceedings of the 37th International Conference on Machine Learning},
    numpages = {10},
    series = {ICML'20},
    address = {Virtual}
}

@inproceedings{Resnet,
  title={Deep residual learning for image recognition},
  author={He, Kaiming and Zhang, Xiangyu and Ren, Shaoqing and Sun, Jian},
  booktitle={Proceedings of the IEEE conference on computer vision and pattern recognition},
  pages={770--778},
  year={2016}
}

@misc{layernorm,
      title={Layer Normalization}, 
      author={Jimmy Lei Ba and Jamie Ryan Kiros and Geoffrey E. Hinton},
      year={2016},
      eprint={1607.06450},
      archivePrefix={arXiv},
      primaryClass={stat.ML},
}

@inproceedings{MFBPR,
    author = {Rendle, Steffen and Freudenthaler, Christoph and Gantner, Zeno and Schmidt-Thieme, Lars},
    title = {BPR: Bayesian personalized ranking from implicit feedback},
    year = {2009},
    publisher = {AUAI Press},
    address = {Arlington, Virginia, USA},
    booktitle = {Proceedings of the Twenty-Fifth Conference on Uncertainty in Artificial Intelligence},
    pages = {452–461},
    numpages = {10},
    series = {UAI '09}
}

@inproceedings{LightGCN,
    author = {He, Xiangnan and Deng, Kuan and Wang, Xiang and Li, Yan and Zhang, YongDong and Wang, Meng},
    title = {LightGCN: Simplifying and Powering Graph Convolution Network for Recommendation},
    year = {2020},
    publisher = {Association for Computing Machinery},
    address = {New York, NY, USA},
    booktitle = {Proceedings of the 43rd International ACM SIGIR Conference on Research and Development in Information Retrieval},
    pages = {639–648},
    numpages = {10},
    series = {SIGIR '20}
}

@article{MOE,
  title={Adaptive mixtures of local experts},
  author={Jacobs, Robert A and Jordan, Michael I and Nowlan, Steven J and Hinton, Geoffrey E},
  journal={Neural computation},
  volume={3},
  number={1},
  pages={79--87},
  year={1991},
  publisher={MIT Press}
}

@inproceedings{Oracle4Rec,
author = {Xia, Jiafeng and Li, Dongsheng and Gu, Hansu and Lu, Tun and Zhang, Peng and Shang, Li and Gu, Ning},
title = {Oracle-guided Dynamic User Preference Modeling for Sequential Recommendation},
year = {2025},
publisher = {Association for Computing Machinery},
address = {New York, NY, USA},
booktitle = {Proceedings of the Eighteenth ACM International Conference on Web Search and Data Mining},
pages = {363–372},
numpages = {10},
series = {WSDM '25}
}

@inproceedings{TrustSVD,
    author = {Guo, Guibing and Zhang, Jie and Yorke-Smith, Neil},
    title = {TrustSVD: collaborative filtering with both the explicit and implicit influence of user trust and of item ratings},
    year = {2015},
    publisher = {AAAI Press},
    booktitle = {Proceedings of the Twenty-Ninth AAAI Conference on Artificial Intelligence},
    address = {New York, NY, USA},
    pages = {123–129},
    numpages = {7},
    series = {AAAI'15}
}

@article{SR_Survey1,
    author = {Jing, Mengyuan and Zhu, Yanmin and Zang, Tianzi and Wang, Ke},
    title = {Contrastive Self-supervised Learning in Recommender Systems: A Survey},
    year = {2023},
    publisher = {Association for Computing Machinery},
    address = {New York, NY, USA},
    volume = {42},
    number = {2},
    journal = {ACM Trans. Inf. Syst.},
    numpages = {39},
}

@article{SR_Survey2,
    author = {Fang, Hui and Zhang, Danning and Shu, Yiheng and Guo, Guibing},
    title = {Deep Learning for Sequential Recommendation: Algorithms, Influential Factors, and Evaluations},
    year = {2020},
    publisher = {Association for Computing Machinery},
    address = {New York, NY, USA},
    volume = {39},
    number = {1},
    journal = {ACM Trans. Inf. Syst.},
    numpages = {42},
}
